%% file: ms.tex
\documentclass[sigconf, anonymous=false, 10pt]{acmart}
\pdfoutput=1
\usepackage{graphicx}
\usepackage{balance}  
\usepackage{subfigure}
\usepackage{color}
\usepackage[ruled, vlined, linesnumbered]{algorithm2e}
\usepackage{multirow}
\usepackage{appendix}

\newtheorem{definition}{Definition}

\newtheorem{example}{Example}

\setcopyright{acmcopyright}
\copyrightyear{2020}
\acmYear{2020}
\acmDOI{10.1145/1122445.1122456}

\def\figw{0.2465}

\acmConference[SIGMOD'20]{SIGMOD'20: 2020 International Conference on Management of Data}{June 14--19, 2020}{Portland, Oregon, USA}
\acmBooktitle{SIGMOD'20: 2020 International Conference on Management of Data,
	June 14--19, 2020, Portland, Oregon, USA}
\acmPrice{15.00}
\acmISBN{978-1-4503-9999-9/18/06}

\begin{document}

\pagestyle{plain}
\pagenumbering{arabic}

\title{Towards Interpretable and Learnable Risk Analysis for Entity Resolution}

\author{Zhaoqiang Chen$^\dagger$, Qun Chen$^\dagger$, Boyi Hou$^\dagger$, Tianyi Duan$^\dagger$, Zhanhuai Li$^\dagger$ and Guoliang Li$^\ddagger$}

\affiliation{\vspace{-0.15in}$^\dagger$School of Computer Science, Northwestern Polytechnical University, Xi'an, China\\
	$^\dagger$Key Laboratory of Big Data Storage and Management, Northwestern Polytechnical University, Ministry of Industry and Information Technology, Xi'an, China\\
	$^\ddagger$Department of Computer Science, Tsinghua University, Beijing, China
}
\email{{chenzhaoqiang@mail., chenbenben@, ntoskrnl@mail., tianyiduan@mail., lizhh@}nwpu.edu.cn}
\email{liguoliang@tsinghua.edu.cn}

\begin{abstract}
Machine-learning-based entity resolution has been widely studied. However, some entity pairs may be mislabeled by machine learning models and existing studies do not study the risk analysis problem -- predicting and interpreting which entity pairs are mislabeled. In this paper, we propose an interpretable and learnable framework for risk analysis, which aims to rank the labeled pairs based on their risks of being mislabeled. We first describe how to automatically generate interpretable risk features, and then present a learnable risk model and its training technique. Finally, we empirically evaluate the performance of the proposed approach on real data. Our extensive experiments have shown that the learning risk model can identify the mislabeled pairs with considerably higher accuracy than the existing alternatives.
\end{abstract}


\maketitle

\input{1-introduction}
\input{2-relatedwork}
\input{3-framework}
\input{4-feature}

\input{5-riskmodel}

\input{6-experiments}

\input{7-conclusion}

\clearpage
\balance
\bibliographystyle{abbrv}
\bibliography{references}
\end{document}

%% file: 1-introduction.tex
\section{Introduction}

\begin{figure}[!t]
\centering
\subfigure{$R_1$}{
\includegraphics[width=1.0\linewidth]{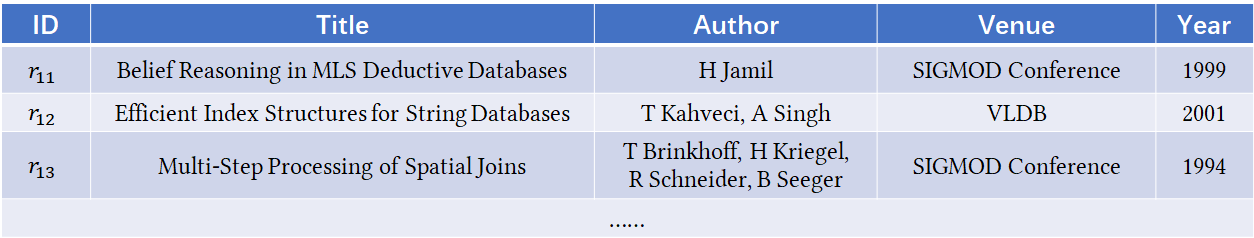}
}
\subfigure{$R_2$}{
\includegraphics[width=1.0\linewidth]{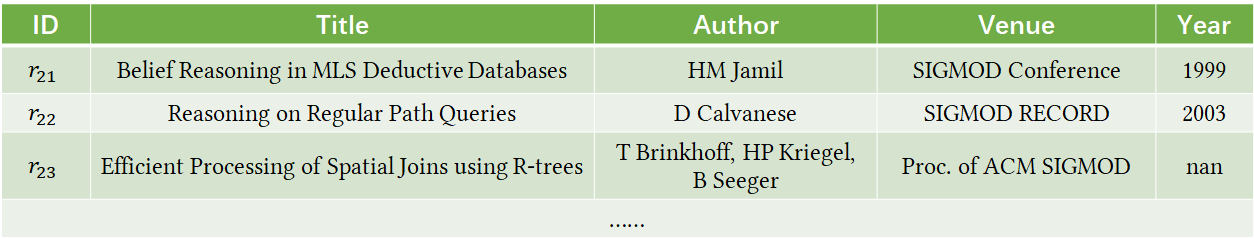}
}
\vspace{-0.1in}
\caption{An ER running example.}
\vspace{-0.1in}
\label{fig:runningexample}
\end{figure}

Entity resolution (ER) aims at identifying the equivalent records that refer to the same real-world entity. Considering the running example in Figure~\ref{fig:runningexample}, ER aims to match the paper records between two tables, $R_1$ and $R_2$. A pair of $<r_{1i},r_{2j}>$, in which $r_{1i}$ and $r_{2j}$ denote a record in $R_1$ and $R_2$ respectively, is called an \emph{equivalent} pair if and only if $r_{1i}$ and $r_{2j}$ refer to the same paper; otherwise, it is called an \emph{inequivalent} pair. In the example, $r_{11}$ and $r_{21}$ are \emph{equivalent} while $r_{11}$ and $r_{22}$ are \emph{inequivalent}. 
		
ER can be taken as a binary classification problem, which aims to label the record pairs as  equivalent/inequivalent. Recently several learning models have been proposed to address the ER problem (most notably among them is deep neural network (DNN)~\cite{joty2018distributed, mudgal2018deep}). Unfortunately, ER can be very challenging in real scenarios due to the prevalence of incomplete and dirty values in the records~\cite{elmagarmid2007duplicate}. Therefore, many record pairs may be mislabeled by a learning model, and the model is hard to interpret such that (1) which pairs are mislabeled, and (2) why these pairs are mislabeled. Thus it is highly desirable to accurately analyze the risk of the labeled pairs returned by a learning model, i.e., ranking the labeled pairs based on their risks of being mislabeled. We call this task {\it risk analysis} for ER, which not only let users know the risks of the labeled results but also evaluate and tune the performance of an ER learning model based on the high-risk pairs. 
	
Although existing learning models can compute a probability indicating the uncertainty of pair status and utilize the probability to quantify the risk, they have several limitations for risk analysis. (1) Not easily interpretable. The existing learning models are hard to interpret. (2) Not learnable. Existing learning models can not be tuned to accurately capture the risk, because they aim to minimize the inconsistency between the predicted result and the ground truth, while risk analysis aims to analyze the risks of labeled results. Therefore, it requires to design a separate model for risk analysis.  

In this paper, we propose a novel framework for interpretable and learnable ER risk analysis. Our work focuses on the learning model that can accurately rank the labeled pairs based on their risks of being mislabeled. Our major contributions are  summarized as follows.

\begin{itemize}
	\item We propose an interpretable and learnable framework for ER risk analysis, which consists of risk feature generation, risk model construction and risk model training. To the best of our knowledge, this is the first interpretable and learnable approach for risk analysis. 
	\item We design a learning model to evaluate the risks of labeled pairs in ER and propose a training algorithm to efficiently tune the model towards a specific workload. 
	\item We present a technical solution of automatic risk feature generation for ER risk analysis.
	\item We empirically evaluate the performance of the proposed solution on real data by a comparative study. Our extensive experiments have shown that it can identify the mislabeled pairs with considerably higher accuracy than existing alternatives. 
\end{itemize}

This paper considers risk analysis as a separate process independent of ER classifier training. Due to uncertainty and non-interpretability of DNN output, it has been well recognized in the AI community~\cite{amodei2016concrete} that risk analysis is a vital issue to AI safety. As a result, separate risk analysis has been actively studied in the literature\cite{zhang2014predicting, hendrycks2016baseline, hendrycks2018deep, jiang2018trust}. Compared with DNN output, our proposed approach, referred to as LearnRisk in this paper, can effectively improve both accuracy and interpretability. Even though we focus on ER in this paper, the proposed approach can be potentially generalized to other classification tasks. On the other hand, it is worthy to point out that the applications of risk analysis are not limited to risk ranking, which can be directly used to reduce required manual cost in machine and human collaboration for high-quality entity resolution~\cite{hou2018r}. It can also be potentially leveraged to optimize the essential processes of classifier training, including active selection of training instances and model training. We discuss how to leverage risk analysis for classifier training in Section~\ref{sec:potential-application}. 

The rest of this paper is organized as follows: Section~\ref{sec:relatedwork} reviews related work. Section~\ref{sec:problem} defines the ER task. Section~\ref{sec:framework} introduces the general framework. Sections~\ref{sec:risk-feature} and \ref{sec:risk-model} present the technical solution for ER, with Section~\ref{sec:risk-feature} focusing on risk feature generation and Section~\ref{sec:risk-model} on risk model. Section~\ref{sec:experiment} presents our empirical evaluation results. Section~\ref{sec:potential-application} discusses the potential applications of risk analysis. Finally, Section~\ref{sec:conclusion} concludes this paper.
 

%% file: 2-relatedwork.tex
\section{Related work} \label{sec:relatedwork}

We discuss related work from the orthogonal perspectives of risk analysis and entity resolution. 
\vspace{0.05in}	

\noindent{\bf Risk Analysis}. Risk analysis has been alternatively called \emph{confidence ranking} or \emph{trust scoring} in the previous literature. There has been a growing interest in the study of risk analysis in recent years \cite{zhang2014predicting, hendrycks2016baseline, hendrycks2018deep, jiang2018trust}. The authors of \cite{hendrycks2016baseline} showed that the output probabilities from softmax distributions, though may be misleading if viewed in isolation, can perform fairly well in detecting mislabeled data on various tasks including computer vision, natural language processing and automatic speech recognition. In their following work~\cite{hendrycks2018deep}, they proposed to fine tune a pre-trained classifier with an auxiliary dataset of outliers and an outlier exposure module to better detect the out-of-distribution data. Alternatively, the authors of \cite{jiang2018trust} employed the metric of relative cluster distance, which compares the distances of an instance to the cluster with the same label and its nearest cluster with a different label, to measure the risk. However, these proposals are not easily interpretable. Moreover, they are not trainable, thus can not be adapted to a specific workload. While the aforementioned work focused on risk analysis for the tasks of image and speech recognition, the authors of r-HUMO~\cite{hou2018r} proposed an approach of risk analysis for the task of ensuring quality guarantees for entity resolution. The proposed approach was however tailored to the problem setting of quality control. Based on the work of r-HUMO, the same team~\cite{chen2018improving} has proposed a similar but more general approach for risk analysis on ER. It first estimates a pair's equivalence probability distribution by Bayesian inference and then uses the metric of Conditional Value at Risk (CVaR) to measure its risk. Unfortunately, the proposed risk model is not trainable either.

Due to the uncertainty of DNN output, there is a research field of confidence calibration~\cite{platt1999probabilistic, guo2017calibration} that seeks to transform a classifier output into a probability estimate representative of true correctness likelihood. Unfortunately, the calibration techniques usually do not change the ranking order of instances as measured by classifier output. Risk analysis however needs to accurately rank the instances by measured risk. Therefore, they cannot serve as reliable risk indicators. Moreover, similar to DNN output, calibrated output also has the interpretability issue. Another research field related to risk analysis is to provide with explaining models for classification results~\cite{ribeiro2016should,baehrens2010explain,amershi2015modeltracker}. However, the field of model explanation mainly put effort on providing interpretable information for human analysis. In contrast, this paper aims to provide an interpretable and learnable framework for automatic quantitative risk measurement.

\vspace{0.05in}
\noindent{\bf Entity Resolution.} ER plays a key role in data integration and has been extensively studied in the literature~\cite{christen2012data,elmagarmid2007duplicate,christophides2015entity}. ER can be automatically performed based on rules~\cite{fan2009reasoning,li2015rule,singh2017generating}, probabilistic theory~\cite{fellegi1969theory,singla2006entity} and machine learning  models~\cite{sarawagi2002interactive,cochinwala2001efficient,christen2008automatic,kouki2017collective}. The progressive paradigm for ER~\cite{whang2013pay, altowim2014progressive,lacoste2013sigma} has also been proposed for the application scenario in which ER should be processed efficiently but does not necessarily require to generate high-quality results. Taking a pay-as-you-go approach, it studied how to maximize quality given a given budget. 

It has been well recognized that automatic entity resolution can be very challenging in real scenarios due to the prevalence of dirty and incomplete values~\cite{elmagarmid2007duplicate}. Therefore, there have been an enduring interest in involving the human in resolution process for improving the performance. Many active learning techniques have been proposed for the task of ER~\cite{sarawagi2002interactive,mozafari2014scaling,arasu2010active,bellare2012active}. It is noteworthy that the pair selection strategies employed by active learning can naturally serve as the metrics for risk analysis. However, their purpose is to improve the overall performance of an under-developed classifier, but not to evaluate the risk of individual pairs being mislabeled by a well-trained classifier. Thus, as the existing learning models for ER, they are not able to accurately capture the risk of individual pairs. In recent years, many researchers ~\cite{mozafari2014scaling, chai2016cost, wang2012crowder, vesdapunt2014crowdsourcing, Gruenheid2012Crowdsourcing, Getoor2012Entity, Chu2015Katara, Firmani2016Online, whang2013question, gokhale2014corleone, wang2015crowd, verroios2017waldo, li2017human, yang2018cost} have also studied how to crowdsource an ER workload. While these researchers addressed the challenges specific to crowdsourcing, we instead focus on risk analysis for ER in this paper. 

%% file: 3-framework.tex
\section{Problem Statement} \label{sec:problem}

Given some records, entity resolution reasons about whether two records are equivalent. Two records are deemed to be equivalent if and only if they correspond to the same real-world entity. We call a pair an {\em equivalent} pair if and only its two records are equivalent; otherwise, it is called an {\em inequivalent} pair. An ER solution labels each pair in a workload as {\em matching} or {\em unmatching}. Note that given an ER workload, a perfect solution would label each equivalent pair as {\em matching} and each inequivalent pair as {\em unmatching}. However, due to the inherent challenge of entity resolution, a model may be prone to mislabeling some pairs.
	
\begin{table}
\centering
\caption{Frequently Used Notations.}
\vspace{-0.1in}
\label{tb:notations}
\begin{tabular}{|l|l|}
\hline
 Notation & Description \\ \hline
 $R$        & a table consisting of records \\ \hline
 $D$        &  an ER workload consisting of record pairs \\ \hline
 $r_i$ & a record in a table \\ \hline
 $d_i$ & a record pair in $D$ \\ \hline
 $S_i$ & a labeling solution for $d_i$ \\ \hline
 $f_i$ & a feature of a pair \\ \hline
 $F_i$ & a feature set \\ \hline
 $D_f$      & the set of pairs with the feature $f$ \\ \hline
\end{tabular}
\vspace{-0.1in}
\end{table}

Given an ER workload of $D$ and a labeling solution of $S$ for $D$, the task of risk analysis is to rank the pairs in $D$ by risk such that the mislabeled pairs can be generally ranked before the correctly labeled pairs. As in previous work~\cite{hendrycks2016baseline}, we employ the metric of Receiver Operating Characteristic (ROC) curve \cite{fawcett2006introduction} to measure the performance of risk analysis. 
	
\begin{figure}
	\centering
	\includegraphics[width=0.6\linewidth]{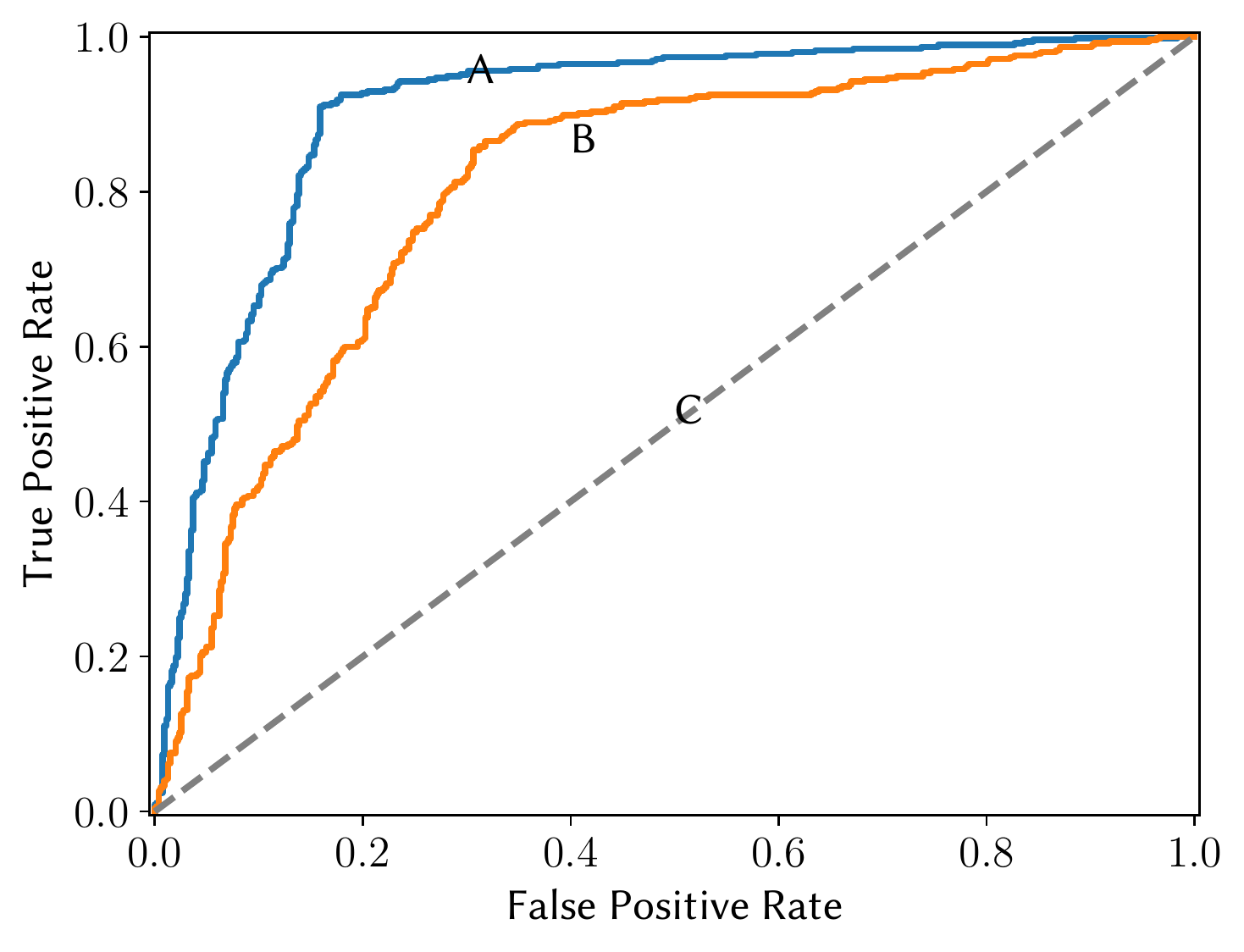}
	\vspace{-0.15in}
	\caption{An examples of ROC curve: C denotes a trivial baseline model, and model A is better than B.}
	\vspace{-0.15in}
	\label{fig:roc-example}
\end{figure}

Let \emph{TP} denote the number of true positives, \emph{FP} the number of false positives, \emph{TN} the number of true negatives and \emph{FN} the number of false negatives. Then, the true positive rate, denoted by \emph{TPR}, is equal to $\frac{TP}{TP+FN}$, and the false positive rate, denoted by \emph{FPR}, is equal to $\frac{FP}{FP+TN}$. Note that in the circumstance of risk analysis, a positive pair refers to a mislabeled pair and a negative pair refers to a correctly labeled pair. 	
ROC curve illustrates \emph{TPR} against \emph{FPR} at different threshold settings. It depicts the relative tradeoff between benefit (true positives, i.e. the mislabeled pairs) and cost (false positives, i.e. the correctly labeled pairs) \cite{fawcett2006introduction}.  By the metric of ROC, Area Under ROC (AUROC) can be deemed as the probability that a risk model assigns higher score to a randomly chosen positive pair than a randomly chosen negative pair \cite{fawcett2006introduction, hendrycks2016baseline}. Therefore, a model with a higher AUROC achieves better quality. Note that a trivial model without discrimination has the AUROC of $50\%$.  An example of ROC curve has been shown in Figure \ref{fig:roc-example}. 
	
\begin{figure*}
	\centering
	\includegraphics[width=0.94\linewidth]{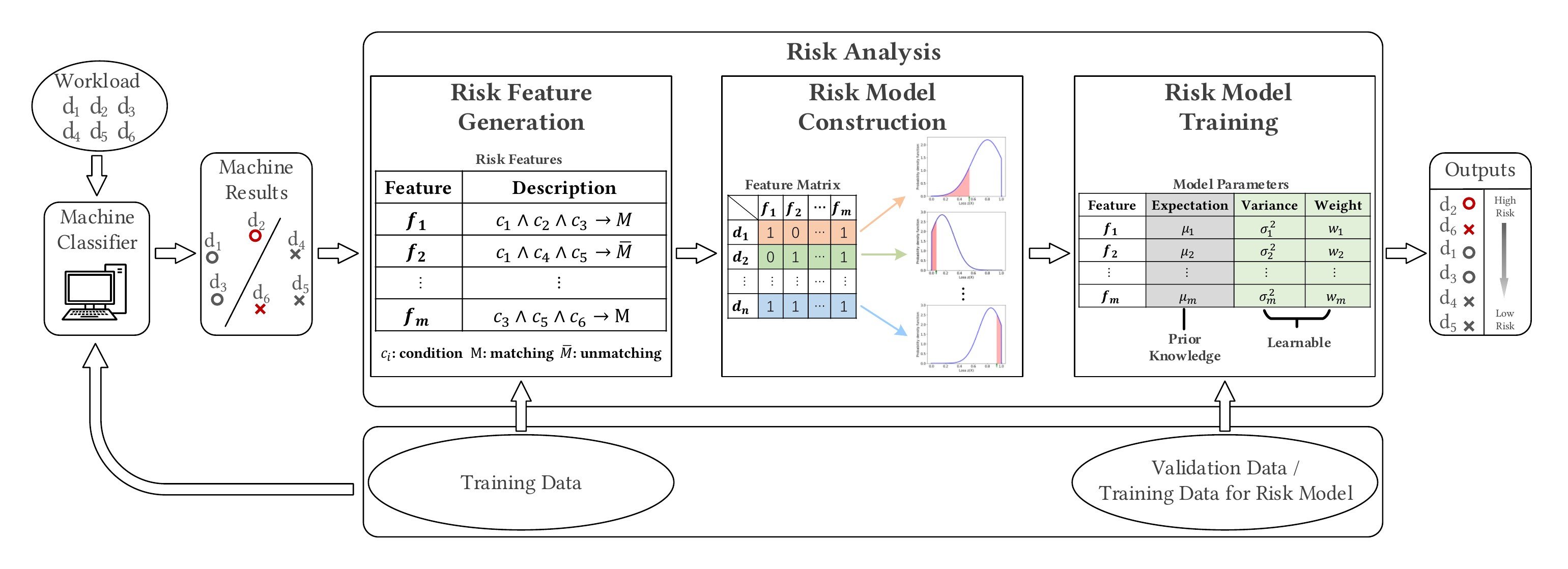}
	\vspace{-0.15in}
	\caption{The framework of risk analysis: the circle and cross symbols represent different classes, and the mislabeled pairs are highlighted by {\color[rgb]{0.7529,0,0} red}.}
	\vspace{-0.1in}
	\label{fig:framework}
\end{figure*}

For presentation simplicity, we summarize the frequently used notations in Table~\ref{tb:notations}. Formally, we define the task of risk analysis as follows:
	
\begin{definition} \label{problemdefinition}
	{\bf [Risk Analysis for ER].} Suppose that an ER workload of $D$ consists of $n$ record pairs, \{$d_1$, $d_2$, $\ldots$, $d_n$\}, and a machine classifier of $S$ labels each pair in $D$ as {\em matching} or {\em unmatching}. The task of risk analysis is to quantitatively measure the risk of each pair in $D$ being mislabeled such that the metric of AUROC is maximized. 
\end{definition}

\section{Risk Analysis Framework} \label{sec:framework}

The general framework of risk analysis, as shown in Figure~\ref{fig:framework}, consists of the following three steps:
\begin{itemize}
   \item {\bf Risk feature generation};
	 \item {\bf Risk model construction};
	 \item {\bf Risk model training};	
\end{itemize}
In the rest of this section, we will elaborate each of the three steps.

\subsection{Risk Feature Generation}  

The framework measures mislabeling risk based on features. To enable interpretable and learnable risk analysis, the risk features should have the following three desirable properties:
\begin{itemize}
\item {\bf Interpretable.} For a risk model to be interpretable, its risk features should be interpretable, or easily understandable to the human.
\item {\bf Discriminating.} For a risk model to be effective, its risk features should be highly discriminating, or indicative of the ground-truth labels of pairs. In contrast, a non-discriminating feature can not serve as an effective risk feature, because it can not provide valuable insights into class status.
\item {\bf High-coverage.}  For a risk model to be learnable, its risk features should represent the common knowledge shared among many pairs. High coverage ensures that the knowledge learned on training data can be effectively transferred to test data. In contrast, a low-coverage feature has limited utility for risk analysis because the knowledge it embodies is not easily transferable. 
\end{itemize}

The challenge of risk feature generation is then how to automatically generate interpretable, discriminating and high-coverage risk features. In the scenario of ER, we observe that rule is the most common form of knowledge that can be easily understood by the human. Therefore, we propose to extract the rules satisfied by a pair as its risk features. 
For instance, in the running example shown in Figure~\ref{fig:runningexample}, if two records have different publication years, it is unlikely that they refer to the same paper. This knowledge can be represented by the rule of
\begin{equation} \label{eq:rule-example}
  r_i[Year]\neq r_j[Year] \rightarrow inequivalent(r_{i}, r_{j}),
\end{equation}	
in which $r_i$ denotes a record and $r_i[Year]$ denotes $r_i$'s value at the attribute of $Year$. 
With this knowledge, a \emph{matching} pair whose two records have different publication years can be reasoned to be at high risk of being mislabeled. We discuss how to automatically generate risk features in Section~\ref{sec:risk-feature}.

\subsection{Risk Model Construction} \label{sec:general-risk-model}
	
\begin{figure}
	\centering
	\includegraphics[width=\linewidth]{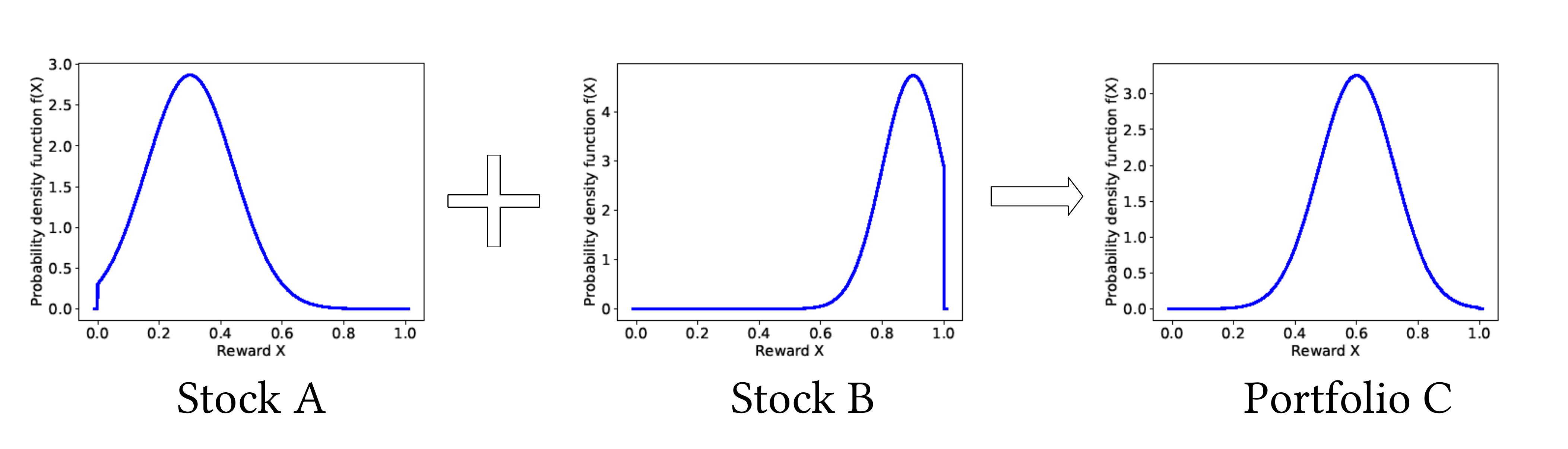}
	\vspace{-0.2in}
	\caption{Estimation of the reward distribution of an investment portfolio: aggregating the reward distributions of its component stocks.}
	\vspace{-0.12in}
	\label{fig:portfolio}
\end{figure}

For each extracted risk feature, the framework models its equivalence probability by a distribution. The step of risk model construction then estimates the equivalence probability distribution of each pair based on its risk features, and quantifies its risk based on the estimated distribution. 
Inspired by the success of risk analysis in investment theory, we propose to construct risk model based on the theory of portfolio investment~\cite{rockafellar2002conditional}. Specifically, the framework considers each risk feature as a stock and each pair as a portfolio of component stocks. It models the distribution of a risk feature (resp. a pair) by the reward distribution of its corresponding stock (resp. portfolio). As shown in Figure~\ref{fig:portfolio}, the reward distribution of a portfolio (resp. a pair) is estimated by aggregating the reward distributions of its component stocks (resp. risk features).  
	
On the task of ER, we take the label of each pair as a random variable, which follows the Bernoulli distribution with the parameter $p$ representing its equivalence probability. Statistically speaking, the parameter $p$ follows the Beta distribution 
$Beta(\alpha,\beta)$~\footnote{https://en.wikipedia.org/wiki/Conjugate\_prior}, where $\alpha$ and $\beta$ denote the shape parameters. If the values of $\alpha$ and $\beta$ are large (>=10), the Beta distribution can be approximated by the Normal distribution~\cite{wise1960normalizing}. Note that the value of ($\alpha + \beta$) can be interpreted as sample size. In the case of ER, sample size is usually large. Therefore, we model the equivalence probability of an ER pair by normal distribution. 

Specifically, the framework denotes the normal distribution of the equivalence probability of a labeled pair by $\mathcal{N}(\mu_i, \sigma_i^2)$, where $\mu_i$ and $\sigma_i^2$ denote its expectation and variance respectively. Since a valid equivalence probability should be between 0 and 1, we transform an inferred normal distribution to a \emph{truncated form} in the range of 0 to 1~\cite{burkardt2014truncated}. Suppose that there are $m$ risk features, \{$f_1$, $f_2$, $\ldots$, $f_m$\}. Let $\mathbf{w}=[w_1, w_2,$ $...,$ $w_m]^T$ denote the feature weights, and $\mathcal{N}(\mu_{f_j},\sigma_{f_j}^2)$ denote the equivalence probability distribution of the feature $f_j$. Accordingly, the expectation vector of $m$ features is denoted by $\mathbf{\mu}_F=$ $[\mu_{f_1},$ $\mu_{f_2},$ $\ldots, \mu_{f_m}]^T$, and the variance vector by $\mathbf{\sigma}^2_F=$ $[\sigma_{f_1}^2,$ $\sigma_{f_2}^2,$ $\ldots, \sigma_{f_m}^2]^T$. For each pair $d_i$, we denote its feature vector by $\mathbf{x}_i=[x_{i1}, x_{i2}, ..., x_{im}]$, where $x_{ij}=1$ if $d_i$ has the $j$th feature, otherwise $x_{ij}=0$. Then, the distribution of $d_i$ is represented by $\mathcal{N}(\mu_i,\sigma_i^2)$, in which 
\begin{equation}
  \mu_i = \mathbf{x}_i (\mathbf{w} \circ \mathbf{\mu}_F), 
\end{equation}
and
\begin{equation}
  \sigma_i^2 = \mathbf{x}_i (\mathbf{w}^2 \circ \mathbf{\sigma}^2_F), 
\end{equation}
where $\circ$ represents the elementwise product. Finally, provided with the distribution of a pair, the framework uses the popular metrics proposed for investment risk measurement (e.g. Value at Risk (VaR)) to quantify its risk of being mislabeled. A risk model would rank all the labeled pairs based on their quantified risks. 

It is noteworthy that the existing approaches based on DNN output use a single value to represent equivalence probability, which corresponds to the expectation in our proposed risk model. However, in investment risk analysis, it has been observed that besides expected return, return fluctuation also plays an important role in risk estimation. The scenario of ER risk analysis is similar in that equivalence probability fluctuation also brings additional uncertainty. Therefore, instead of using a single value, our proposed risk model uses normal distribution to more accurately capture the uncertainty of label status. As shown in investment risk analysis~\cite{tardivo2002value}, the metric of VaR, which estimates the highest probability of an instance being mislabeled in the majority of cases, is quite effective at capturing fluctuation risk. Therefore, we use the metric of VaR in this paper. However, it is noteworthy that other metrics can be similarly used in the framework. We discuss how to construct the risk model in Section~\ref{sec:risk-model}.
	
\subsection{Risk Model Training} 

Provided with a risk model, the final step is to tune the model towards a specific workload such that it can flexibly reflect the characteristics of the workload. Similar to traditional classifier model training, risk model training primarily involves parameter tuning. Since the framework aims to rank the labeled pairs by the risk of being mislabeled, it requires a solution of learning to rank~\cite{burges2005learning}.
	
Specifically, the proposed risk model has three types of parameters, the weight of risk feature ($w_i$), the expectation ($\mu_i$) and variance ($\sigma_i^2$) of risk feature distribution. Typically, the framework considers the expectation of feature distribution as prior knowledge and estimates it based on the labeled data used for classifier training. It instead tunes the parameters of $w_i$ and $\sigma_i^2$ for optimal risk ranking. Desirably, the training data for risk model should be directly selected from a target workload. In practice, the machine learning solutions usually require additional labeled data for classifier validation. They can be leveraged for risk model training.  We discuss how to train risk model in Section~\ref{sec:risk-model}.
	

%% file: 4-feature.tex
\section{Risk Feature Generation} \label{sec:risk-feature}

The framework uses rules to represent risk features. 
We note that decision tree \cite{cochinwala2001efficient} or its variation, random forest \cite{gokhale2014corleone}, has been widely used to generate the ER rules. Aiming to directly label a workload, the existing techniques of decision tree construction would generate a limited number of labeling rules. Provided with a labeling rule, if a pair satisfies the condition specified at Left-Hand Side (LHS), it would be labeled as the class specified at Right-Hand Side (RHS); otherwise, it is implied that the pair would be labeled as a different class. For instance, consider the labeling rule specified as follows
\begin{equation} \label{eq:two-sided-rule-example}
  sim(r_i[Title],r_j[Title])>0.9 \rightarrow equivalent(r_{i}, r_{j}),
\end{equation}	
in which $r_i[Title]$ denotes $r_i$'s value at the attribute of $Title$, and $sim(\cdot,\cdot)$ measures the similarity between two attribute values. The rule reasons that if the similarity between two records' attribute values at $Title$ is larger than 0.9, they are equivalent; \emph{otherwise, they are inequivalent.}  

In contrast, the rule of risk feature is one-sided. One-sidedness means that if a pair satisfies the condition specified at LHS, it is very likely that it belongs to the class specified at RHS; however, nothing needs to be implied if a pair does not satisfy the condition. For instance, consider the rule specified in Eq.~\ref{eq:rule-example}, which states that if two paper records have different publication years, it is very unlikely that they refer to the same paper. It is not a good labeling rule because even though two papers share the same publication year, it can not be said with high confidence that they refer to the same paper. However, it can serve as an effective risk feature. 

In the rest of this section, we first describe how to design basic metrics based on attribute values for ER and then present an approach to automatically generate the interpretable, discriminating and high-coverage risk features based on the basic input metrics.

\subsection{Basic Metric Design}

\begin{figure}[!t]
	\centering
	\includegraphics[width=0.95\linewidth]{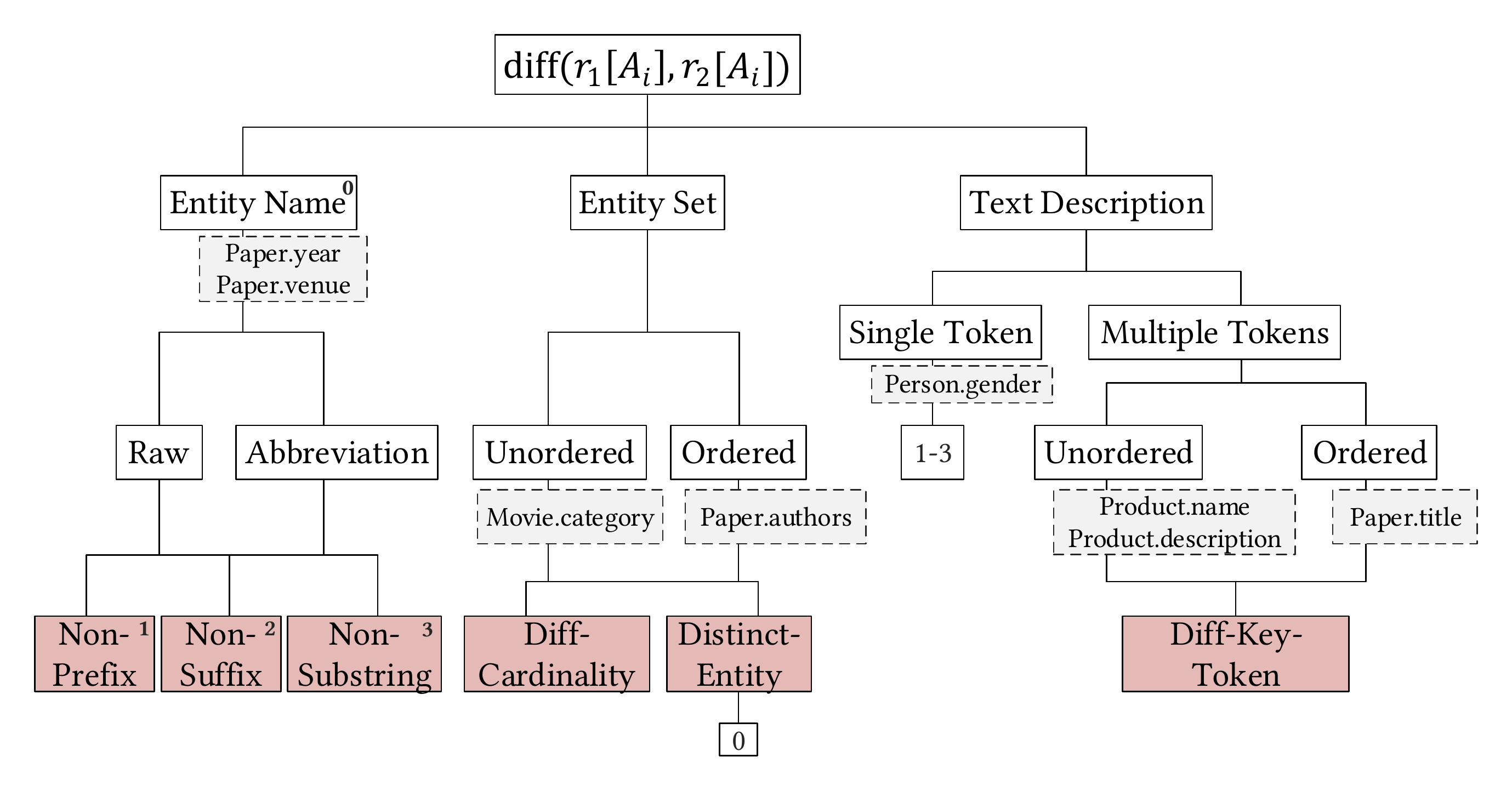}
	\vspace{-0.15in}
	\caption{Summary of the difference metrics defined on string attributes.}
	\vspace{-0.12in}
	\label{fig:attr-comparison}
\end{figure}

Similar to the existing rule-based ER solutions~\cite{singh2017generating}, we specify rules by the comparison operations between attribute values. We note that a wide variety of similarity metrics have been proposed for different types of attribute values (e.g., string and numerical)~\cite{christen2012data}. Since the similarity metrics focus on the common part between two values, they are usually effective as the indicators of equivalence status. Since low similarity usually means low equivalence probability, similarity metrics can also be used to indicate inequivalence status. However, for the reasoning of inequivalence status, the metrics that directly capture the difference between two values may be more effective. For instance, in the running example as shown in Figure~\ref{fig:runningexample}, if two paper records have different numbers of authors, it is very likely that they refer to different papers even though they may share most of the authors. Consider another scenario where a paper record contains a very specific token in its title, which however does not occur in the title of another record. It is very likely that these two records refer to different papers even though their titles may appear highly similar.
	
Therefore, besides similarity metrics, we also define \emph{difference metrics}, which are denoted by \emph{diff}$(r_1[A_i],$ $r_2[A_i])$, to directly capture the difference between two attribute values. Since the comparisons between numerical values are straightforward (e.g., $>$, $<$, $=$ and $\neq$), as usual we focus on string values in this paper. Even though there exist many off-the-shelf string similarity metrics, there are very few difference metrics defined in the literature as far as we know. 
Therefore, in the rest of this subsection, we present the difference metrics defined on string values.  	
We categorize them by the type of string values, which includes entity name, set of entity names, and text description:

\begin{itemize}	

\item \underline{\bf Entity name}. To specifically capture the difference between two entity names, we define the difference metrics of \emph{non-substring} and \emph{abbr-non-substring}, \emph{non-prefix} and \emph{abbr-non-prefix}, \emph{non-suffix} and \emph{abbr-non-suffix}. The metric of \emph{non-substring} (resp. \emph{non-prefix} and \emph{non-sufffix}) indicates whether a value is a substring (resp. prefix and suffix) of another value. Similarly, the metric of \emph{abbr-non-substring} (resp. \emph{abbr-non-prefix} and \emph{abbr-non-suffix}) indicates whether the first-letter abbreviation of a value is a substring (resp. prefix and suffix) of another value abbreviation. It can be expected that a metric value of 1 for these metrics usually indicates that two entity names represent different entities.

\vspace{0.1in}
\item \underline{\bf Entity set}. To specifically capture the difference between two entity sets, we propose the difference metrics of \emph{diff-cardinality} and \emph{distinct-entity}. The metric of \emph{diff-cardinality} indicates whether the number of entity names in two sets are different, while the metric of \emph{distinct-entity} counts the number of distinct entity names, which exist in only one of two sets.

\begin{example}
		Suppose that two different papers have the author list of $s_1$ = ``T Brinkhoff, H Kriegel, R Schneider, B Seeger'' and $s_2$ = ``T Brinkhoff, H Kriegel, B Seeger'' respectively, in which each author is identified by the comma splitter. The similarity metric of \emph{entity-based JaccardIndex} is measured at 0.75, which tends to label them as an equivalent pair; while the difference metric of \emph{distinct-entity} is 1, which captures the different author ``R Schneider''. It can be observed that in this example, \emph{distinct-entity} is a more effective indicator of inequivalence status than \emph{entity-based JaccardIndex}.
\end{example}
	
\item \underline{\bf Text description}. An attribute value of text description consists of one or multiple tokens. Unlike entity name, a value of text description usually has no abbreviation, and contains a long text (e.g., the attribute of \emph{title} in a paper table). To specifically capture the difference between two text values, we propose the metric of \emph{diff-key-token}. The metric of \emph{diff-key-token} indicates the number of the key or discriminating tokens contained by one and only one record in a pair. A discriminating token can effectively identify an entity, can thus serve as an effective indicator of inequivalence status. 

\end{itemize}

We have summarized the difference metrics in the hierarchical structure as shown in Figure~\ref{fig:attr-comparison}. With the help of hierarchical guidance, we can easily design proper difference metrics for various string attributes. 

\subsection{Rule Generation}

	In constructing a traditional \emph{two-sided} decision tree, a partition operation divides an ER pair set of $D$ into two separate subsets, $D_L$ and $D_R$, both of which consist of mostly equivalent or inequivalent pairs.
	In contrast, the process of rule generation for risk analysis is one-sided. At each iteration, it only needs to extract a subset of pairs $D_i$ from $D$ such that most of the pairs in $D_i$ have the same class status. The remaining pairs in the subset of $D'$=$D$-$D_i$ can instead have mixed labels. The extraction operation is supposed to be iteratively applied on $D'$.
	
  In the construction of two-sided decision tree, the label purity of a partition operation $o$ on a pair set $D$ is usually measured by the metric of \emph{Gini index}~\cite{breiman2017classification}, which is defined by
\begin{equation}
	G(D,o) = \frac{|D_L|}{|D|}G(D_L) + \frac{|D_R|}{|D|}G(D_R),
\label{eq:gini-index}
\end{equation}  
where $G(D_L)$ and $G(D_R)$ denote the \emph{Gini} values of the subsets. The \emph{Gini} value of a pair set is defined by
\begin{equation}
G(D_*) = 1 - t_M^2 - t_U^2,
\label{eq:gini-value}
\end{equation}
where $t_M$ (resp. $t_U$) denotes the proportion of real matches in a subset $D_*$ (resp. real unmatches in $D_*$). It can be observed that the metric of \emph{Gini} value measures the \emph{impurity} of a subset. An optimal partition operation would result in the minimum value of $G(D, o)$. 

   Similarly, in the construction of one-sided decision tree, we design the one-sided Gini index as follows
\begin{equation}
\hat{G}(D, o) = min(\frac{\lambda}{|D_L|} + (1 - \lambda)G(D_L), \frac{\lambda}{|D_R|} + (1 - \lambda)G(D_R)),
\label{eq:one-sided-Gini}
\end{equation}
where $\lambda$ is the weight parameter to balance the influence of size and impurity. 
A large $\lambda$ means we prefer set size over set purity. Since the extracted subset needs to be highly discriminating, we suggest to set the value of $\lambda$ at low (e.g. $0.2$ in our implementation). A partition operation would generate two subsets with the minimum value of one-sided Gini index. By this setting, the construction process achieves the purpose that each partition would produce a highly pure leaf node regardless of the purity of the other one. 

\vspace{0.1in}
\hspace{-0.15in}{\bf Algorithm.}  We have sketched the procedure of automatic rule generation in Algorithm~\ref{alg:rule-generation}. 
	Given a pair set of $D$, it exhaustively applies each of the basic metrics on $D$. For each basic metric, the algorithm chooses the best condition value on the metric, which results in the minimum Gini value, to extract a subset of mostly equivalent or inequivalent pairs. 
	Note that each partition operation on $D$ would generate a separate decision tree. The partition operation is iteratively invoked on the subset consisting of the remaining pairs until either the remaining pairs satisfies a pre-specified purity threshold, or the depth of decision tree reaches a pre-specified threshold. We have implemented the algorithm based on the open-sourced project of two-sided decision tree construction\footnote{https://scikit-learn.org/stable/modules/tree.html}. Due to the imbalance problem in the ER task (i.e., there usually exist much more inequivalent pairs than equivalent ones), for the generation of matching rules, the procedure sets a large class weight (e.g. 1000 in our implementation) to the matching instances. However, the generated matching rules are finally filtered without class weighting. To ensure high coverage of risk feature, we also set a lower threshold on the sheer size of any extracted subset (e.g. 5 in our implementation).
The algorithm would generate a forest of decision trees, in which each leaf corresponds to a rule or risk feature. 	
	
\begin{figure}
	\centering
	\includegraphics[width=\linewidth]{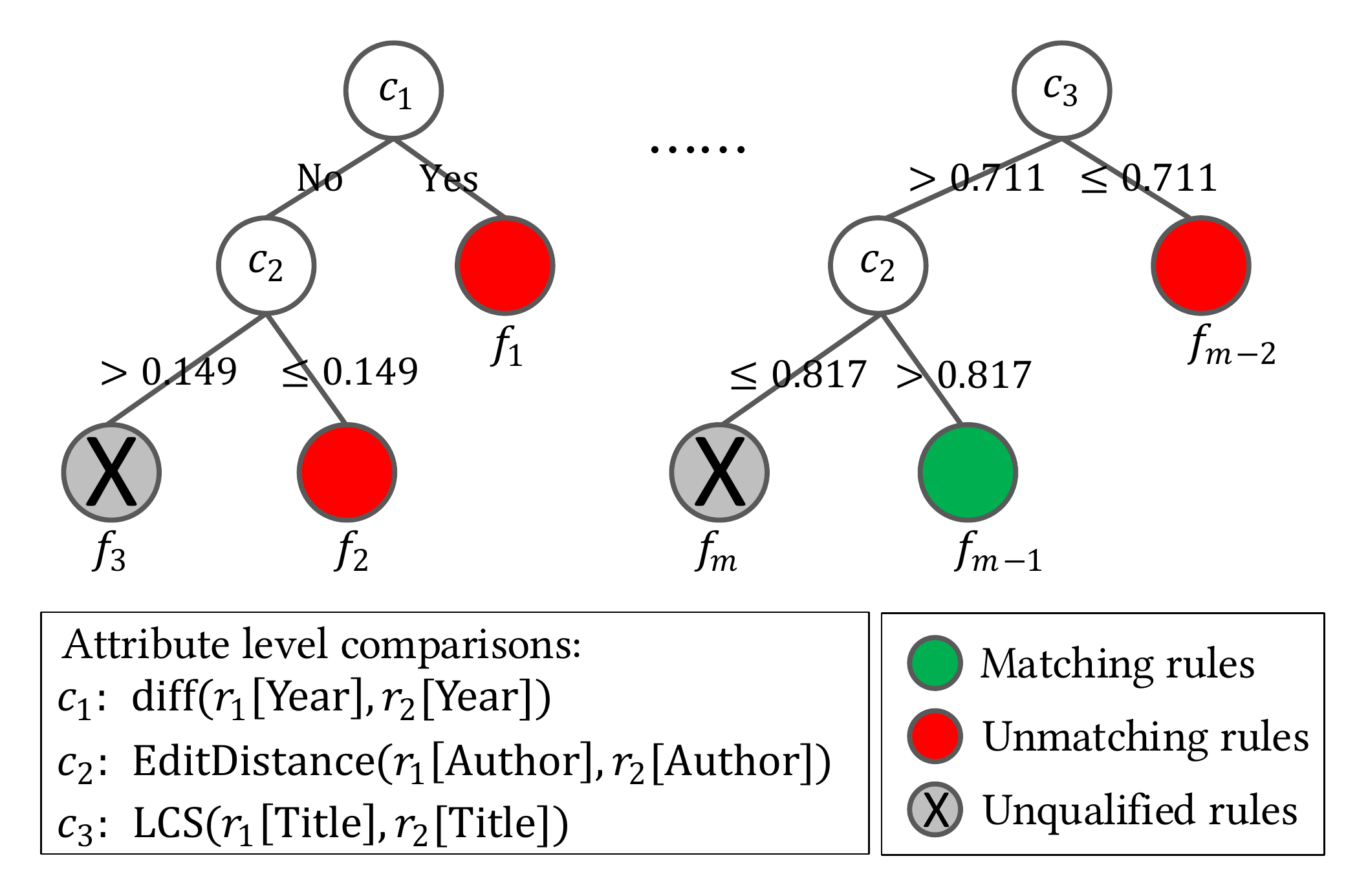}
	\vspace{-0.2in}
	\caption{Illustration of automatic rule generation: each path from the root to a leaf in one-sided decision trees corresponds to a rule. }
	\vspace{-0.1in}
	\label{fig:decisiontree}
\end{figure}

The example rules generated by constructing one-sided decision trees have been shown in Figure \ref{fig:decisiontree}, in which $f_1$ and $f_2$ are both valid rules, and $f_3$ is however not valid because its purity does not exceed the threshold. In Algorithm~\ref{alg:rule-generation}, the input basic metrics are the similarity and difference metrics specifically designed for ER. However, it can be observed that the general approach for risk feature generation are applicable to other classification tasks. Provided with another classification task, we only need to replace the basic similarity and difference metrics; the algorithm similarly applies.

\vspace{0.1in}
\hspace{-0.15in}{\bf Complexity.}
Let $n$ denote the size of training data, $m$ the number of basic metrics, and $h$ the pre-specified depth of decision trees. The total time complexity of rule generation can be represented by $O(h(2m)^h \cdot nlogn)$. It is worthy to point out that the number of basic metrics ($m$) is usually limited (e.g. dozens); to ensure interpretability, the maximum depth of decision tree ($h$) is usually set to a small value (e.g., $h \leq 4$). Therefore, the algorithm for automatic rule generation can be executed efficiently. Our evaluation in Subsection~\ref{sec:scalability} has also shown that it scales well with the size of training data. 

\begin{algorithm}[!t]
	\caption{Risk Feature Generation}
	\label{alg:rule-generation}
	\KwIn{A set of labeled data $D_{t}$\; \qquad \quad \ \  Impurity threshold $\tau$\; \qquad \quad \ \  Tree depth $h$\;}
	\KwOut{Risk features $F$.}
	$F \gets \{ \}$\;
	$current\_depth \gets 0$\;
	$tree \gets null$\;
	$ConstructTree(D_{t}, current\_depth, tree, F)$\;
	Remove redundant rules in F\;
	return $F$\;
	\BlankLine
	\BlankLine
	\setcounter{AlgoLine}{0}
	\SetKwProg{myproc}{Procedure}{}{}
	\myproc{ConstructTree($D_c, depth, tree, F$)}
	{
		\eIf{$depth >= h$}
		{
			$rules \gets$ traverse the $tree$, select the leaves whose impurities do not exceed $\tau$\;
			$F \gets F + rules$\;
		}
		{
			\For{each attribute $A_i \in \mathbb{A}$ \label{alg1Line:attr}} 
			{
				\For{$w_{class} \in \{weight_0, weight_1\}$ \label{alg1Line:classweight}} 
				{
					$o_* \gets argmin_{o\in A_i}\hat{G}(D_c, o, w_{class})$\;
					Split $D_c$ into $D_L$ and $D_R$ based on $o_*$\;
					Add the information of $o_*, D_L, D_R$ to $tree$\;
					Calculate the impurities of $D_L$ and $D_R$, denoted by $\tau_L$ and $\tau_R$\;
					$\tau_{min} \gets min(\tau_L, \tau_R)$\;
					$\tau_{max} \gets max(\tau_L, \tau_R)$\;
					\eIf{$\tau_{min} >= \tau$ or $\tau_{max} < \tau$ \label{alg1Line:imstart}}
					{
						$rules \gets$ traverse the $tree$, select the leaves whose impurities do not exceed $\tau$\;
						$F \gets F + rules$\; \label{alg1Line:imend}
					}
					{
						\eIf{$\tau_L > \tau_R$}
						{\mbox{$ConstructTree(D_L, depth+1, tree, F)$}}
						{\mbox{$ConstructTree(D_R, depth+1, tree, F)$}}	
					}
					\mbox{Remove $o_*, D_L, D_R$ information from $tree$\;}
				}
			}	
		}
	}
\end{algorithm}

%% file: 5-riskmodel.tex
\section{Risk Model} \label{sec:risk-model}

Given an ER workload of $D$, we denote the equivalence probability of each pair $d_i$ in $D$ by a random variable, $p_i$. The framework models $p_i$ by a normal distribution, $\mathcal{N}(\mu_i, \sigma_i^2)$, where $\mu_i$ and $\sigma_i^2$ denote its expectation and variance respectively. As mentioned in Section~\ref{sec:general-risk-model}, the distribution of $p_i$ is supposed to be estimated by aggregating the distributions of risk features. In this section, we first introduce the metric of Value at Risk (VaR) to quantify the risk of a pair, and then present the techniques for risk model training.

\subsection{Risk Metric}

As in portfolio investment~\cite{artzner1999coherent}, we employ the popular metric of Value at Risk (VaR) to measure the risk of a pair being mislabeled by the machine. Given a confidence level of $\theta$, VaR represents the maximum loss after excluding all worse outcomes whose combined probability is at most $1-\theta$. Formally, given the \emph{loss} function $z(X)\in L^p(\mathcal{F})$ of a portfolio $X$ and $\theta$, the metric of VaR is defined as follows:
\begin{equation}
	VaR_{\theta}(X) = inf\{z_*: P(z(X)\geq z_*)\leq (1-\theta)\}.
\end{equation}

\begin{figure}
	\centering
	\includegraphics[width=0.6\linewidth]{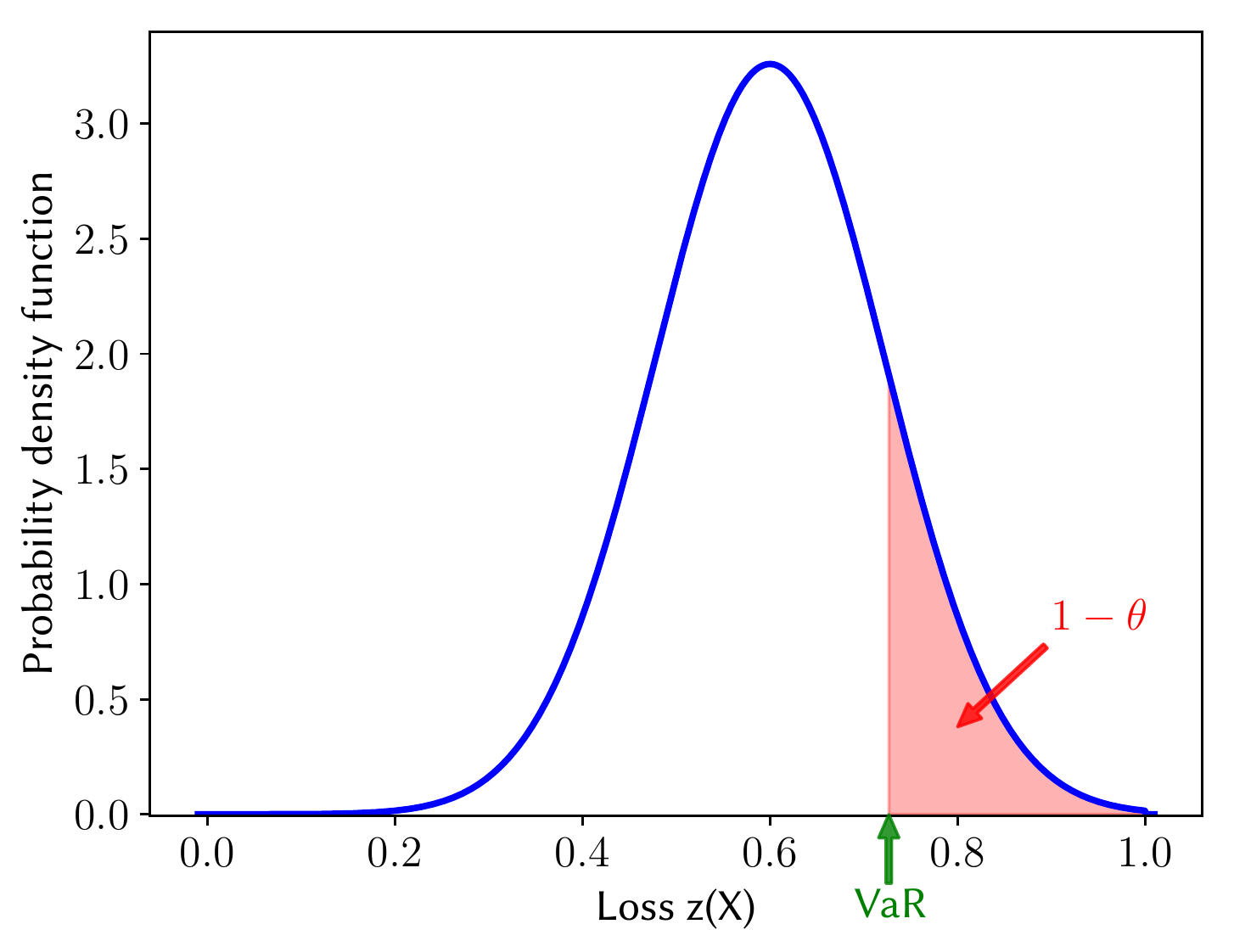}
	\vspace{-0.15in}
	\caption{Visualization of the VaR risk metric.}
	\vspace{-0.15in}
	\label{fig:VaR}
\end{figure}

Given a pair $d_i$, we denote its equivalence probability by $p_i$, and the inverse of its cumulative distribution function by $F_i^{-1}(\cdot)$, i.e., the quantile function. If $d_i$ is labeled as \emph{unmatching} by the machine, its probability of being mislabeled by the machine is equal to $p_i$. Accordingly, its worst-case loss corresponds to the case that $p_i$ is maximal. Therefore, given the confidence level of $\theta$, the VaR of $d_i$ is the maximum value of $z=p_i$ after excluding the $1-\theta$ worse cases where $p_i$ is from $F_i^{-1}(\theta)$ to $+\infty$. Formally, the VaR risk of a pair $d_i$ with the machine label of \emph{unmatching} can be estimated by
\begin{equation}
VaR_{\theta}(d_i) = F_i^{-1}(\theta; \mu_i, \sigma_i^2).
\end{equation}

Similarly, if $d_i$ is labeled by the machine as \emph{matching}, its VaR risk of a pair $d_i$ can be estimated by
\begin{equation}
VaR_{\theta}(d_i) = 1 - F_i^{-1}(1-\theta; \mu_i, \sigma_i^2).
\end{equation}

We have visualized the risk metric of VaR by a pair example with the machine label of \emph{unmatching} in Figure \ref{fig:VaR}, in which the X axis represents the loss and the Y axis represents the probability density function. In Figure \ref{fig:VaR}, the area of red zone is $1-\theta$, which corresponds to the probability of a loss being larger than 0.757. In this case, the VaR value is equal to 0.757.

\subsection{Model Training} 
In this subsection, we detail how to train a risk model. We first describe the parameter setting, and then present the objective loss function and the technique of parameter optimization. 

\subsubsection{Parameter Setting} 

\begin{figure}
	\centering
	\includegraphics[width=0.6\linewidth]{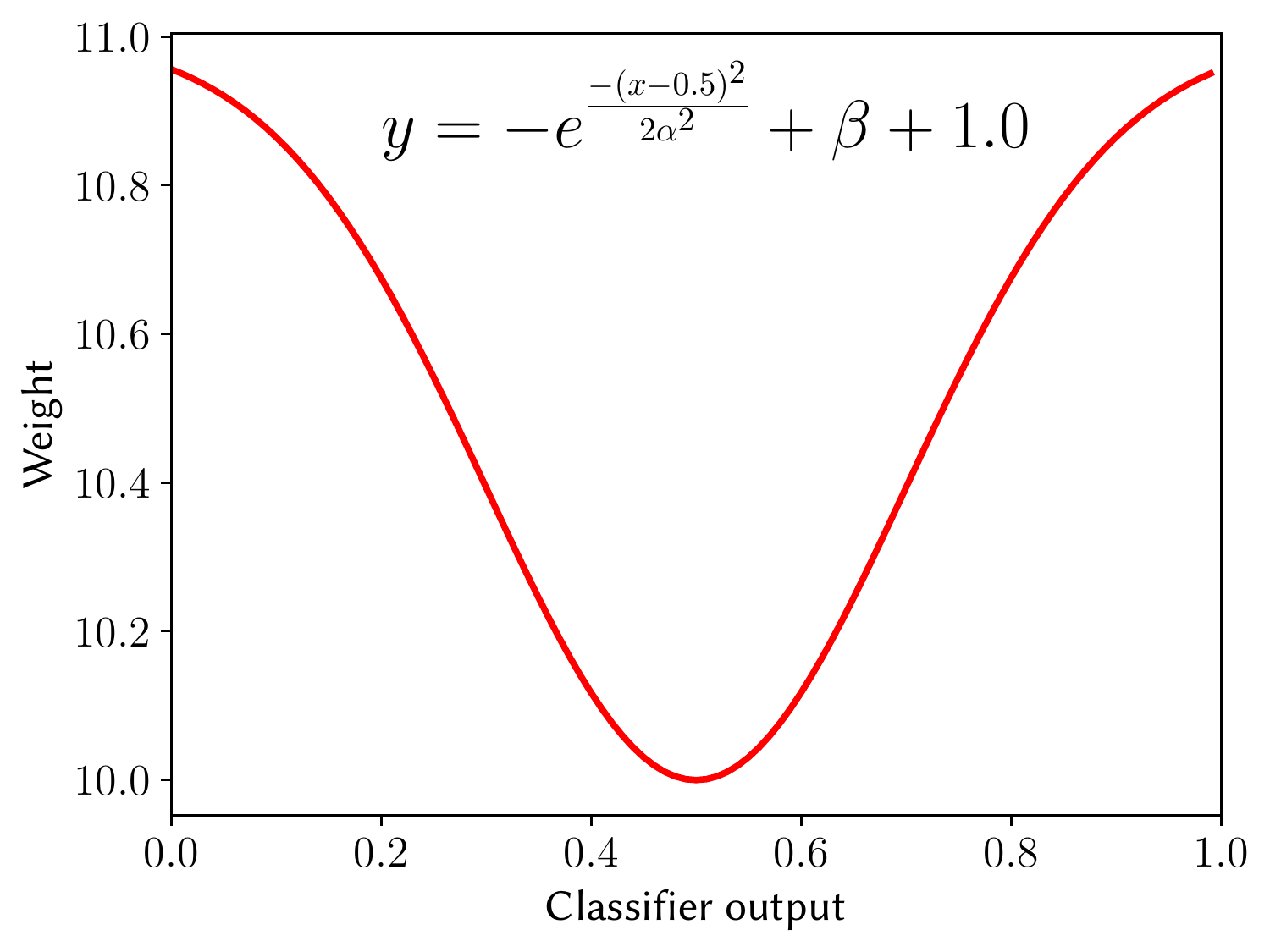}
	\vspace{-0.15in}
	\caption{An example of the influence function, where $\alpha=0.2$ and $\beta=10$.}
	\vspace{-0.15in}
	\label{fig:influence-function}
\end{figure}

  It has been well recognized that although machine classifiers, if used separately, are not able to accurately measure the risk of labeled pairs, their outputs can usually provide with valuable hints on the risk. Specifically, a probability output close to 0 or 1 usually means that the target pair is at low risk of being mislabeled, while an output close to the ambiguous value of 0.5 usually indicates a high risk. Therefore, besides the rules generated by one-sided random forest, the risk model also incorporates classifier output as one of the risk features. 
Furthermore, it is desirable that the weight of a classifier output increases with the extremeness of its value.
 Therefore, we model the influence of classifier output on risk measurement by the function as follows 
\begin{equation}
f_w(x) = -e^{\frac{-(x-0.5)^2}{2\alpha^2}} + \beta + 1.0,
\end{equation}
where $x$ denotes the classifier output, $\alpha$ and $\beta$ are the shape parameters that need to be learned. An example of the influence function has been presented in Figure~\ref{fig:influence-function}. It can be observed that as expected, the influence increases with the extremeness of the output value. With regard to classifier output, the process of model training only needs to learn two parameters, $\alpha$ and $\beta$, instead of the $n$ parameters for each distinct value. 

The model considers the expectations of risk feature distributions as prior knowledge and estimates them by statistical analysis on the labeled data used to train a classifier. Specifically, given a risk feature, $f_i$, suppose that there are totally $n_i$ pair instances with the feature $f_i$ in the classifier training data, and $m_i$ instances among them have the label of \emph{matching}. Then, the expectation of $f_i$ is estimated at $\frac{m_i}{n_i}\times 100\%$.

The feature weights and the variances of risk features instead need to be learned. It can be observed that in general, larger variance means higher risk. We use the metric of \emph{Relative Standard Deviation} (RSD) to measure the uncertainty of feature distribution. Formally, RSD is defined by
\begin{equation}
	RSD=\sigma_i / \mu_i,
\end{equation}	
in which $\mu_i$ and $\sigma_i$ denote the expectation and standard deviation of a feature distribution respectively. The process of model learning needs to learn a value of $RSD$ for each risk feature. For the risk feature of classifier output, we split the pairs into multiple subsets, each of which contains similar classifier outputs. The process of model learning only needs to learn a value of $RSD$ for each subset. 

\subsubsection{Loss Function}

The primary purpose of risk analysis is to prioritize risky pairs such that the mislabeled pair can be ranked higher than the correctly labeled ones. Therefore, we employ the technique of learning to rank~\cite{burges2005learning} to optimize model parameters such that the risk model can best reflect the characteristics of a target workload. 
	
Given any two pairs, $d_i$ and $d_j$, suppose that their risks of being mislabeled are measured at $\gamma_i$ and $\gamma_j$ respectively.  We denote $d_i \rhd d_j$ if and only if $d_i$ is ranked higher than $d_j$ in terms of risk, or $\gamma_i > \gamma_j$. Let $p_{ij}$ denote the posterior probability of $d_i \rhd d_j$. As in ~\cite{burges2005learning}, we use the logistic function to map the risk value to the posterior probability by
\begin{equation} 
p_{ij} = \frac{e^{(\gamma_i - \gamma_j)}}{1 + e^{(\gamma_i - \gamma_j)}}.
\label{eq:posterior-probability}
\end{equation}
According to Eq.~\ref{eq:posterior-probability}, the posterior probability increases with the quantitative risk difference between $d_i$ and $d_j$, or the value of ($\gamma_i - \gamma_j$).

  We set the desired target value of $p_{ij}$ to
\begin{equation} 
\bar{p}_{ij} = 0.5 * (1 + \hat{g}_i - \hat{g}_j),
\label{eq:target-value}
\end{equation}
where $\hat{g}_i, \hat{g}_j \in \{0, 1\}$ are the risk labels of pairs. If a pair $d_i$ is mislabeled, $\hat{g}_i=1$; otherwise, $\hat{g}_i=0$.
According to Eq.~\ref{eq:target-value}, if $d_i$ is mislabeled and $d_j$ is correctly labeled, the value of $\bar{p}_{ij}$ is 1; if instead $d_i$ is correctly labeled and $d_j$ is mislabeled, the value of $\bar{p}_{ij}$ is 0. The general objective of parameter optimization is to minimize the difference between $p_{ij}$ and $\bar{p}_{ij}$. In other words, if $d_i$ is mislabeled and $d_j$ is correctly labeled, the optimization process would maximize the value of ($\gamma_i - \gamma_j$). It corresponds to maximally ranking $d_i$ before $d_j$ in terms of risk. Therefore, our value settings of $p_{ij}$ and $\bar{p}_{ij}$ exactly maximize the metric of AUROC as described in Section~\ref{sec:problem}.

 Formally, given a risk training dataset $D_{\gamma}$, as in~\cite{burges2005learning}, we define the optimization objective by the cross-entropy loss function of
\begin{equation}
L(D_{\gamma}) = \sum_{d_i, d_j \in D_{\gamma}}{-\bar{p}_{ij}\cdot log(p_{ij}) - (1 - \bar{p}_{ij}) \cdot log(1 - p_{ij})}, 
\end{equation} 
in which the value of $L(D_{\gamma})$ increases with the value difference between $p_{ij}$ and $\bar{p}_{ij}$ as desired.

\subsubsection{Parameter Optimization}

We have implemented the process of parameter optimization based on the platform of TensorFlow~\cite{199317}. We employed the widely used technique of \emph{Gradient Descent} to find the parameters that minimize the loss function. Since the loss function $L(D_{\gamma})$ is a composition of differentiable functions (e.g., $log(x)$ and $e^x$), hence $L(D_{\gamma})$ is also differentiable. Given a learning rate $\epsilon$, the parameter $w_i$ is iteratively updated as follows
\begin{equation}
	w_i^{(j+1)} = w_i^{(j)} - \epsilon\cdot\frac{\partial L(D_{\gamma})}{\partial w_i}.
\end{equation} 
Similarly, the parameter $\sigma_i$ is iteratively updated by
\begin{equation}
\sigma_i^{(j+1)} = \sigma_i^{(j)} - \epsilon\cdot\frac{\partial L(D_{\gamma})}{\partial \sigma_i}.
\end{equation}

We set the learning rate $\epsilon$ to 0.001 in our implementation. To alleviate the overfitting problem, we also add a combination of $L_1$ and $L_2$ regularization to the objective loss function.

%% file: 6-experiments.tex
\section{Experiments} \label{sec:experiment}

In this section, we empirically evaluate the performance of the proposed solution, which is denoted by \emph{LearnRisk}, by a comparative study. To the best of our knowledge, no learnable risk model has been proposed in the literature. Therefore, we have compared the proposed solution with the following non-learnable alternatives:

\begin{itemize}

\item \emph{Baseline} \cite{hendrycks2016baseline}. The baseline technique simply measures the risk of a pair by the ambiguity of its classifier output. The pairs with more ambiguous (close to 0.5) outputs (equivalence probabilities) are considered to be at higher risk compared to those with the more extreme outputs. 

\item \emph{Uncertainty} \cite{mozafari2014scaling}: In active learning for ER, the authors of \cite{mozafari2014scaling} proposed to first train multiple classifiers based on different training sets generated by bootstrapping, and then use their predictions to estimate the equivalence probability of a pair. With the equivalence probability estimated at $p$, the risk is measured by the uncertainty score of $p(1-p)$.

\item \emph{TrustScore} \cite{jiang2018trust}: The authors of \cite{jiang2018trust} proposed a clustering-based solution for risk analysis. Firstly, it builds separate clusters to represent each class based on the training data. For a test data $d_i$, let $\rho_Y$ denote its distance to the cluster whose label is the same as $d_i$'s, and $\rho_N$ denote its distance to the nearest cluster whose label is different from $d_i$'s. Then, the \emph{TrustScore} of $d_i$ is estimated by $\frac{\rho_N}{\rho_Y}$. By \emph{TrustScore}, the closer a test data is to its predicted class cluster, the lower risk it has.

\item \emph{StaticRisk} \cite{chen2018improving}: The authors of \cite{chen2018improving} proposed to take the equivalence probability provided by machine classifier as the prior expectation and use the human-labeled pairs as samples to estimate the posterior expectation and variance by the Bayesian inference. Not learnable, \emph{StaticRisk} employed the metric of conditional value at risk to measure pair risk.

\end{itemize}
	
	Besides the aforementioned techniques for risk analysis, we also compare \emph{LearnRisk} with the rule-based solution for ER, \emph{Holoclean}~\cite{rekatsinas2017holoclean}. Note that the existing rule-based solutions for ER aim to label data, while \emph{LearnRisk} ranks the risk of labeled results. They are solving different problems. We however extend Holoclean to support the problem of risk analysis.
	
	The rest of this section is organized as follows: Subsection~\ref{sec:experimentalsetup} describes experimental setup. Subsection~\ref{sec:comparison} evaluates the performance of various risk analysis techniques. Subsection~\ref{sec:holoclean-comparison} compares \emph{LearnRisk} with \emph{Holoclean}. Subsection~\ref{sec:sensitivity} evaluates the performance sensitivity of \emph{LearnRisk} w.r.t the size of risk training data. 
Finally, Subsection~\ref{sec:scalability} evaluates the scalability of \emph{LearnRisk}.	

\vspace{-0.1in}
\subsection{Experimental Setup} \label{sec:experimentalsetup}

   We have used four real datasets in the empirical evaluation, whose details are presented as follows: 
\begin{itemize}
	\item DBLP-Scholar\footnote{\mbox{https://dbs.uni-leipzig.de/file/DBLP-Scholar.zip}} (denoted by DS): the experiments match 2616 DBLP entries from DBLP publications with 64263 Scholar entries from Google Scholar.
	
	\item Abt-Buy\footnote{\mbox{https://dbs.uni-leipzig.de/file/Abt-Buy.zip}} (denoted by AB): the experiments match 1081 product entries from Abt.com with 1092 product entries from Buy.com. 
	
	\item Amazon-Google\footnote{\mbox{https://dbs.uni-leipzig.de/file/Amazon-GoogleProducts.zip}} (denoted by AG): the experiments match 1363 product entities from Amazon and 3226 product entities from Google product search service. Unlike the AB dataset, the AG products are mainly software. 
	
	\item Songs\footnote{\mbox{http://pages.cs.wisc.edu/\~anhai/data/falcon\_data/songs/}} (denoted by SG): the experiments match the entries within a table, which consists of around one million song records. 
\end{itemize}

On all the datasets, we use the blocking technique to filter the pairs deemed unlikely to match. Due to large size of the original SG dataset, we randomly select 5 percent of the records from it as our test workload. The statistics of the four test datasets are presented in Table~\ref{tab:datasets}. 

\begin{table}
	\caption{The statistics of datasets.}
	\vspace{-0.1in}
	\centering
	\label{tab:datasets}
	\begin{tabular}{|c|c|c|c|}
		\hline
		Dataset & Size & \# Matches & \# Attributes \\
		\hline
		DS & 41416 & 5073 & 4 \\
		\hline
		AB & 52191 & 904 & 3 \\
		\hline
		AG & 13049 & 1150 & 4 \\
		\hline
		SG & 144946 & 6842 & 7 \\
		\hline
	\end{tabular}
	\vspace{-0.1in}
\end{table}

We have used DeepMatcher~\cite{mudgal2018deep}, a state-of-the-art deep learning solution for ER, as the machine classifier. For comparative study, as required by DeepMatcher, we split each test dataset into three parts by a pre-specified ratio (e.g. 3:2:5), which specifies the proportions of training, validation and test data respectively. In the comparative study, we have evaluated the performance of different techniques under the circumstances with various ratio settings of training, validation and test data. As expected, our experiments showed that the performance of DeepMatcher generally increases with the proportion of classifier training data. 

For \emph{Uncertainty}, we train 20 deep learning models for each test dataset. For \emph{TrustScore}, the input features are the summaries of attribute similarities, which are represented by 300-dimensions vectors in the deep learning model. For risk feature generation of \emph{LearnRisk}, we have designed 19 basic metrics on the attribute values in the DS workload, which include 8 \emph{diff}$(\cdot)$ metrics; totally 9 basic metrics on attribute values in the AB workload, which include 2 \emph{diff}$(\cdot)$ metric; totally 11 basic metrics on the attribute values in the AG workload, which include 2 \emph{diff}$(\cdot)$ metrics; totally 14 basic metrics in the SG dataset, which include 5 \emph{diff}$(\cdot)$ metrics. In the evaluation of \emph{LearnRisk}, we use the assigned validation data for risk model training. We set the confidence of the VaR risk model at $0.9$ and the number of epochs for parameter optimization at 1000. 

\subsection{Comparative Evaluation} \label{sec:comparison}

\begin{figure*}
	\centering
	\subfigure[DS(1:2:7)]
	{\includegraphics[width=\figw\linewidth]{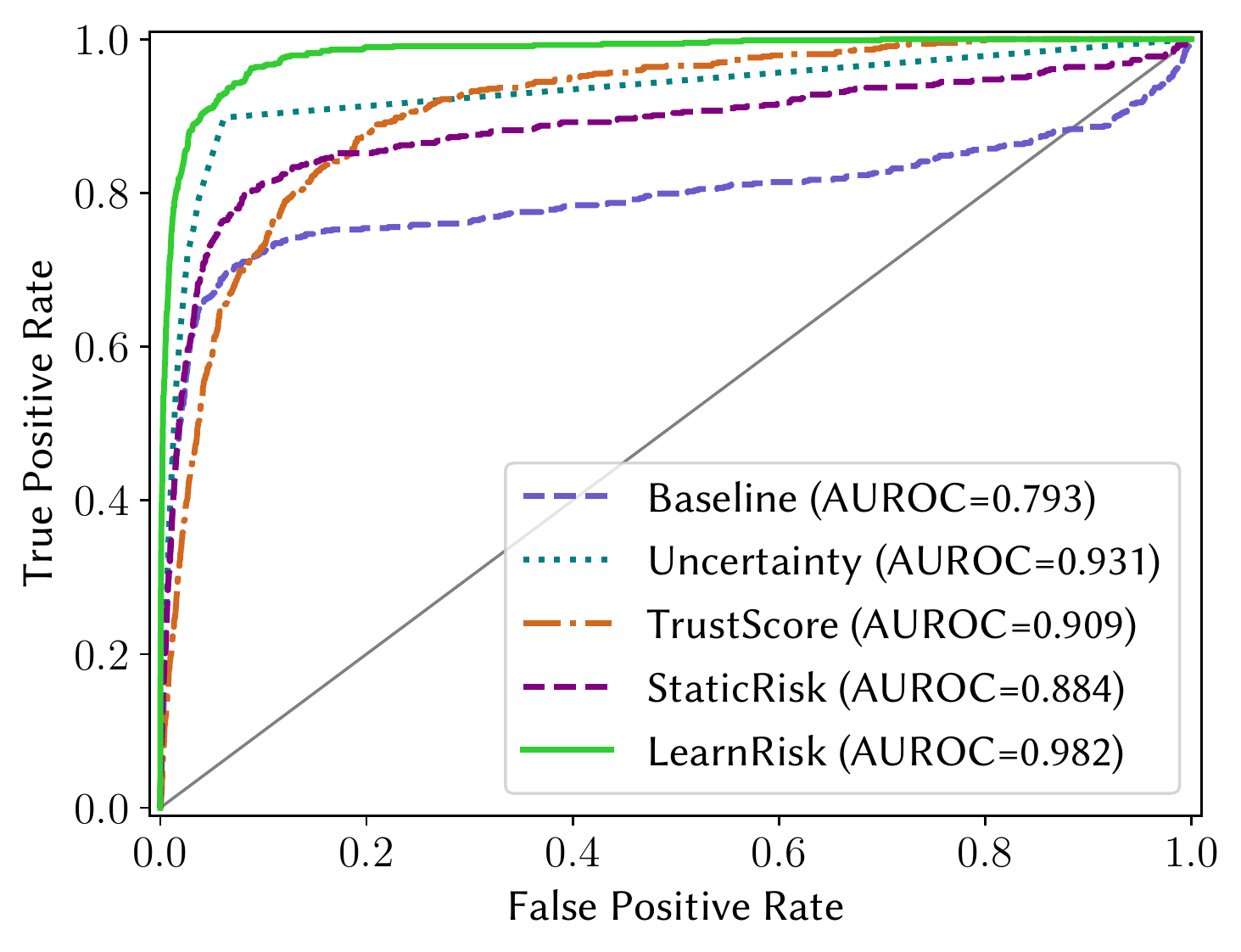}}
	\subfigure[DS(2:2:6)]
	{\includegraphics[width=\figw\linewidth]{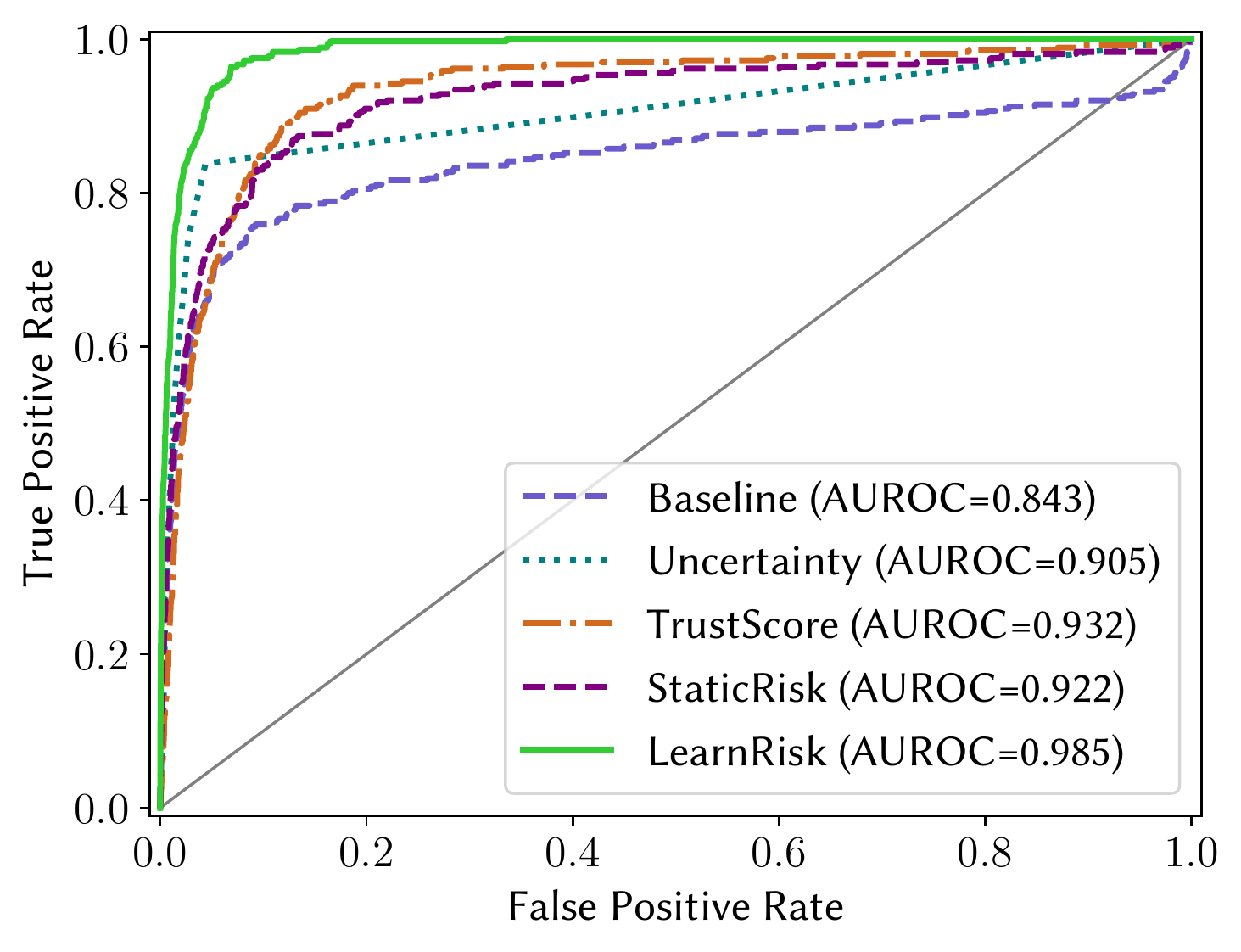}}
	\subfigure[DS(3:2:5)]
	{\includegraphics[width=\figw\linewidth]{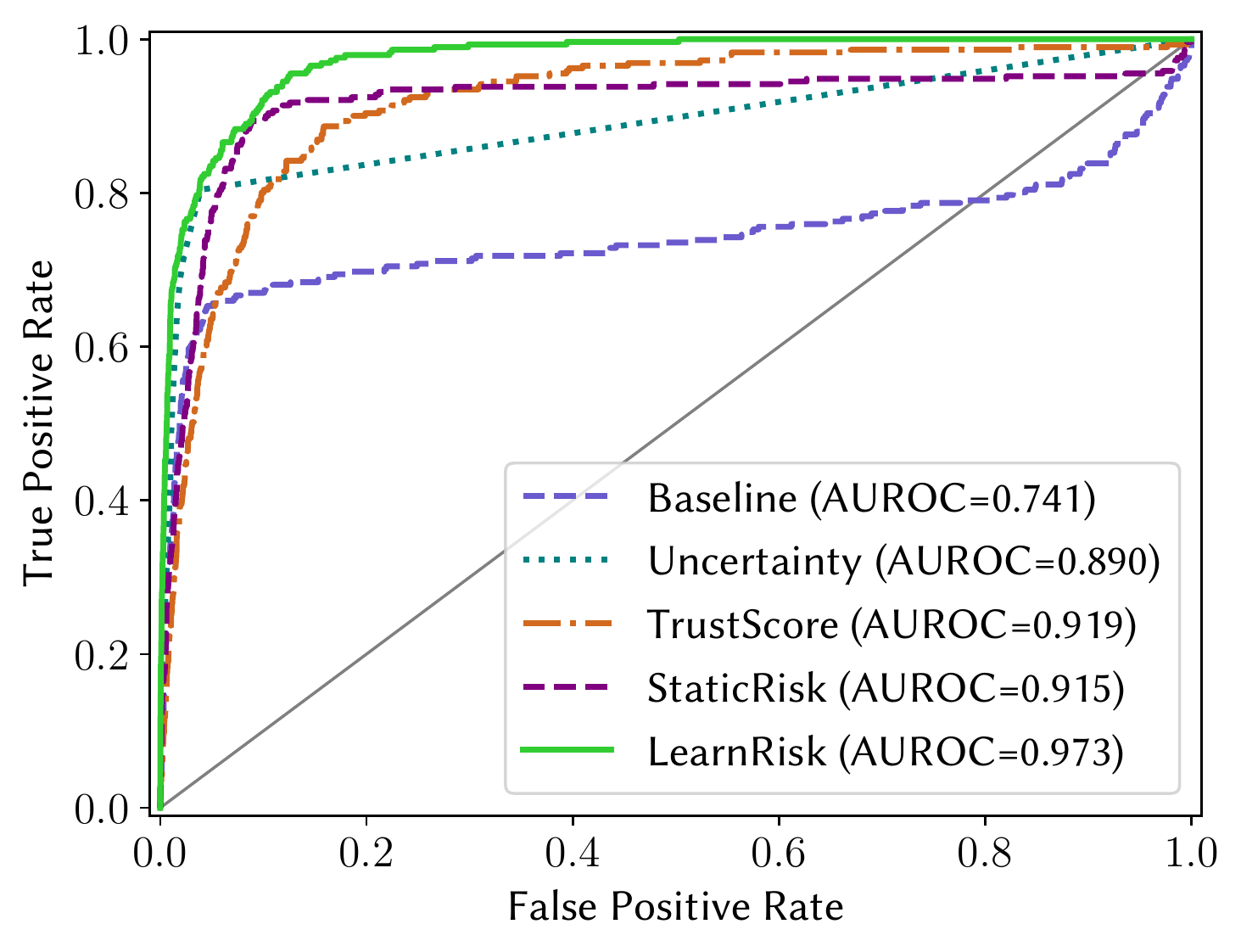}}
	\subfigure[AB(1:2:7)]
	{\includegraphics[width=\figw\linewidth]{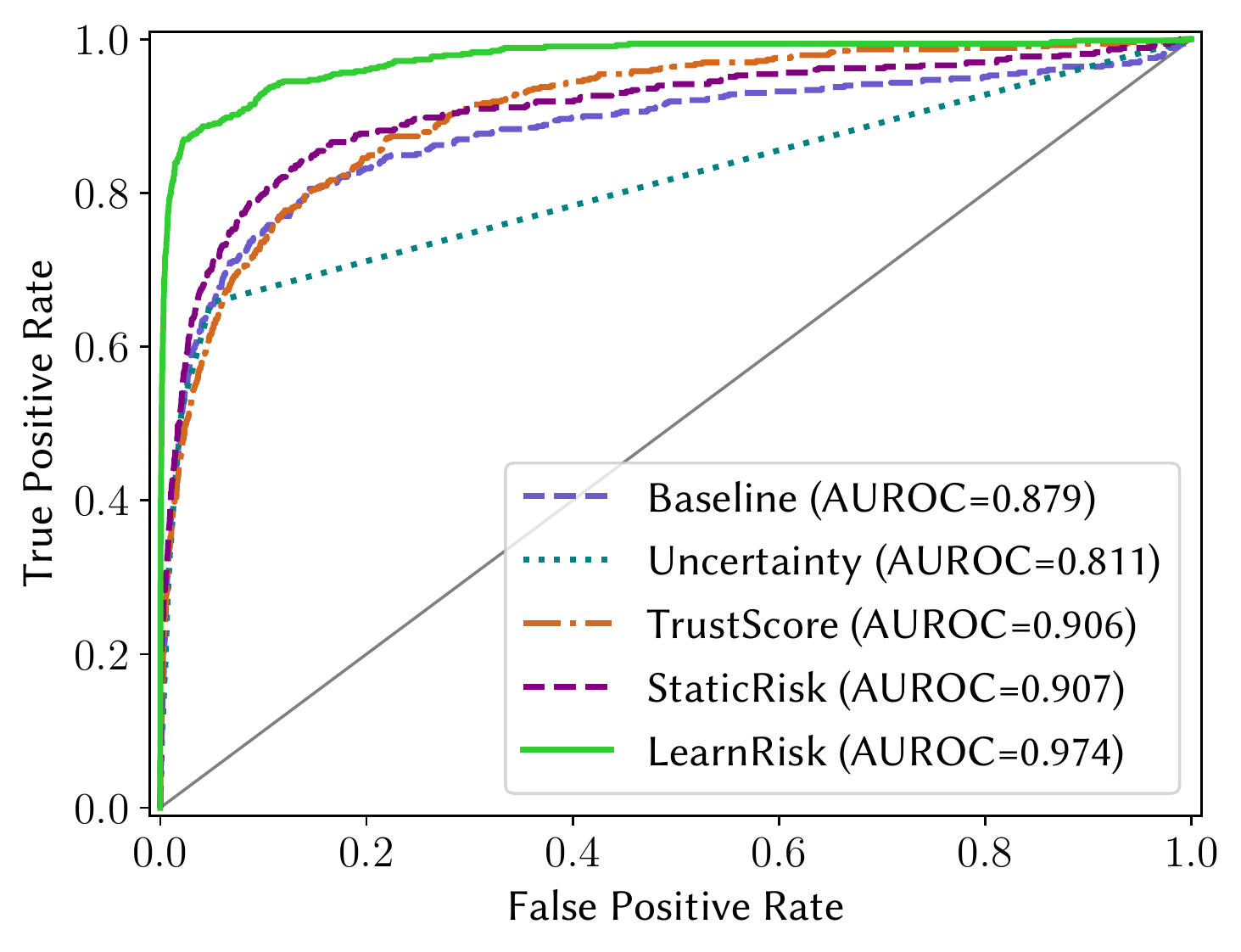}}
	\subfigure[AB(2:2:6)]
	{\includegraphics[width=\figw\linewidth]{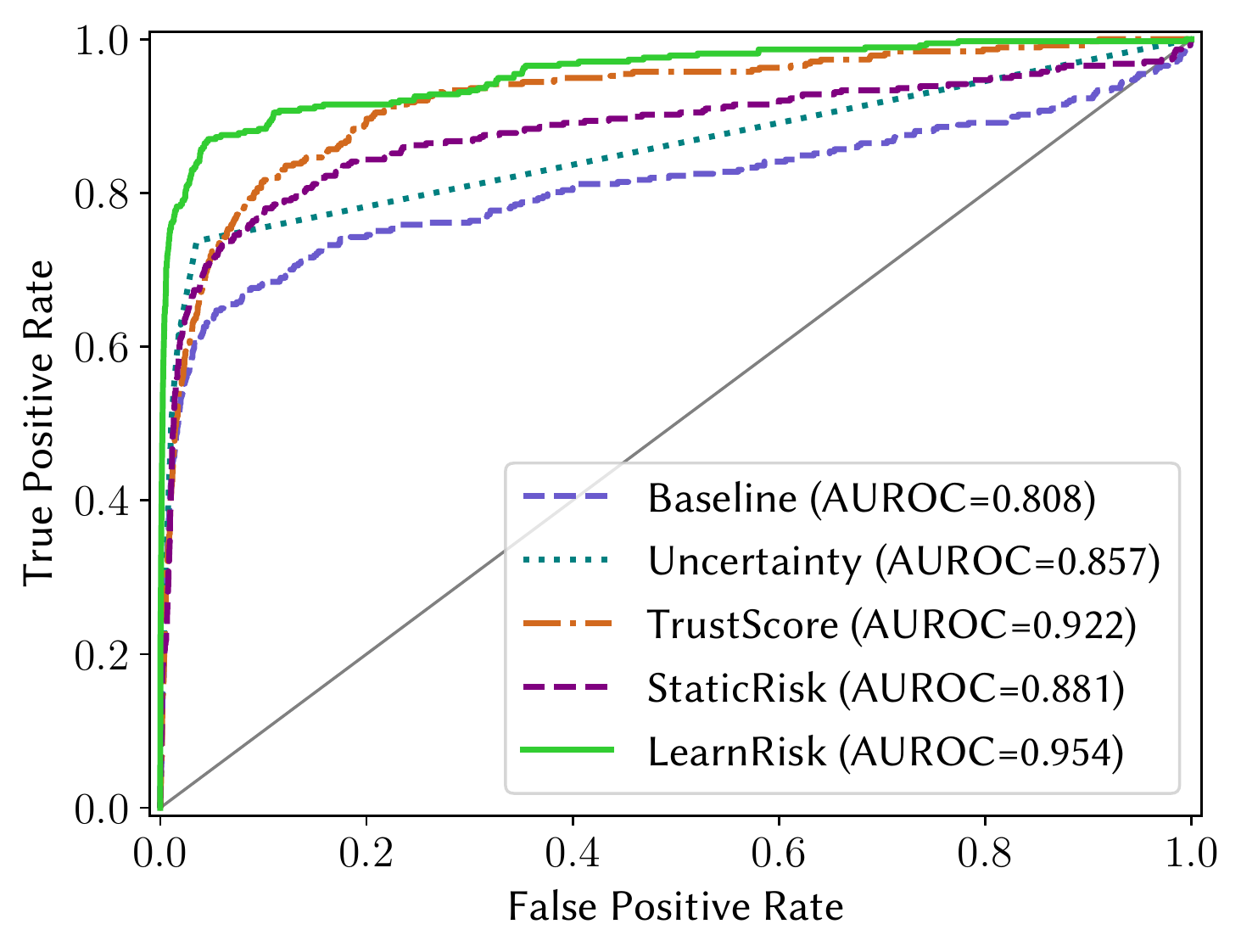}}
	\subfigure[AB(3:2:5)]
	{\includegraphics[width=\figw\linewidth]{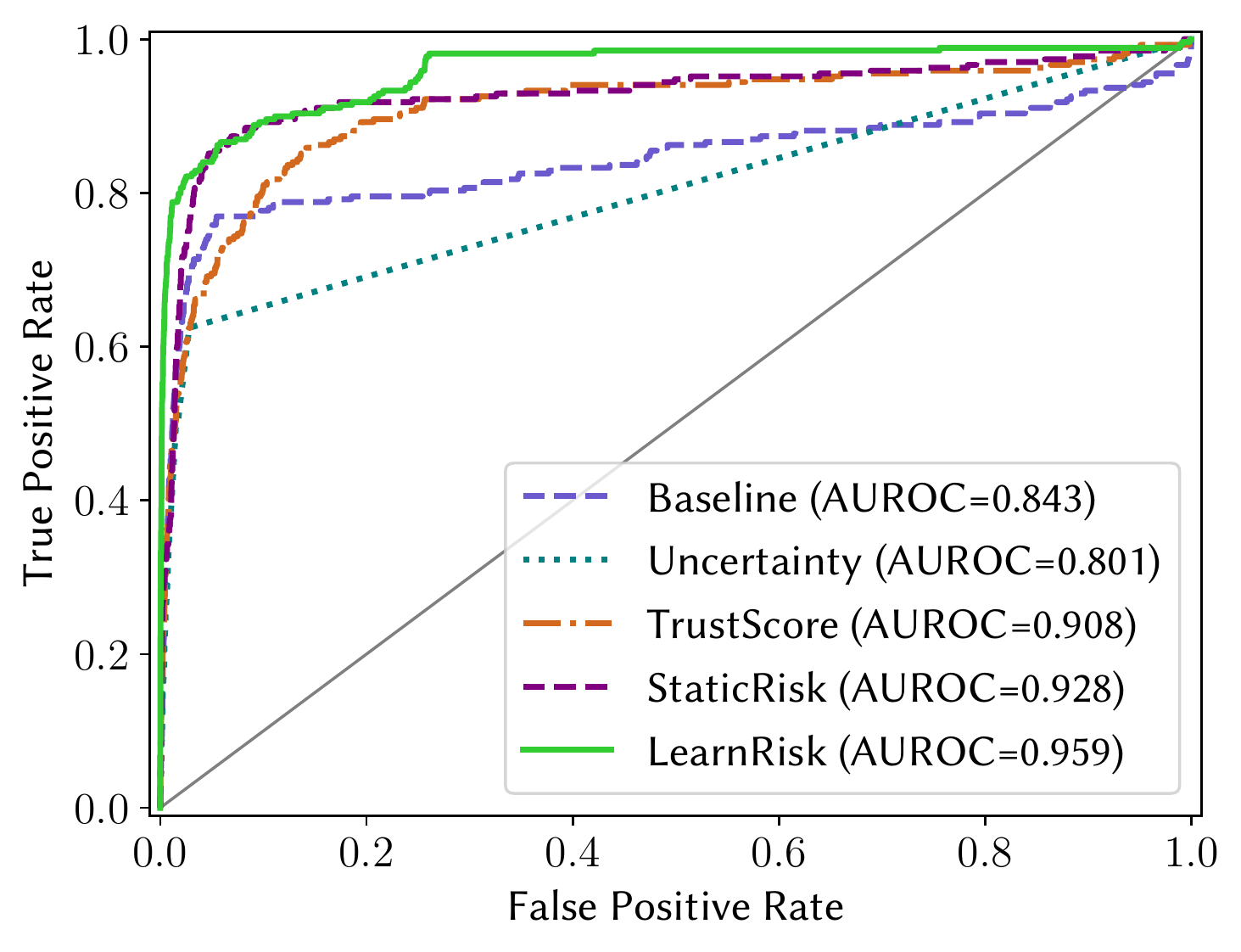}}
	\subfigure[AG(1:2:7)]
	{\includegraphics[width=\figw\linewidth]{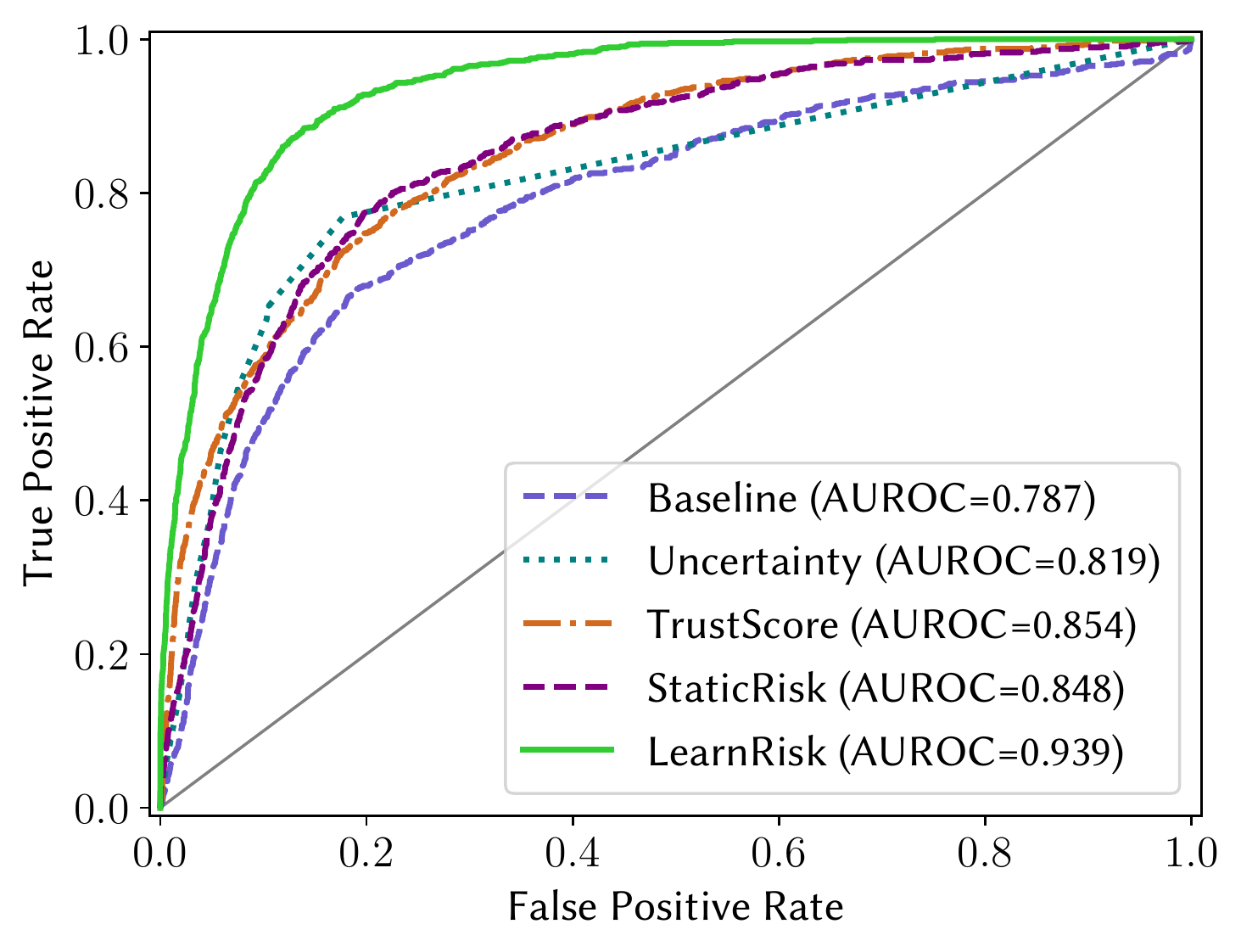}}
	\subfigure[AG(2:2:6)]
	{\includegraphics[width=\figw\linewidth]{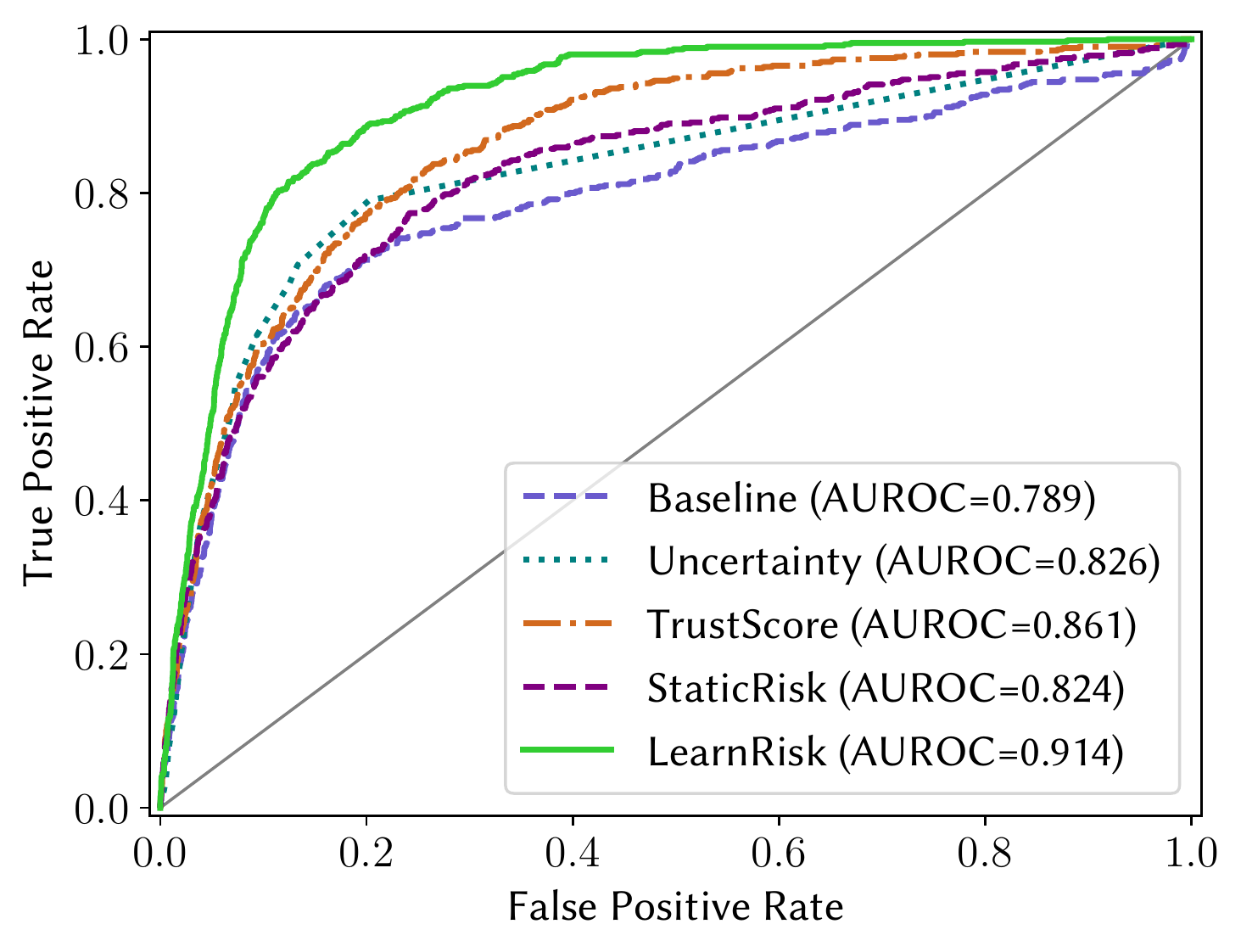}}
	\subfigure[AG(3:2:5)]
	{\includegraphics[width=\figw\linewidth]{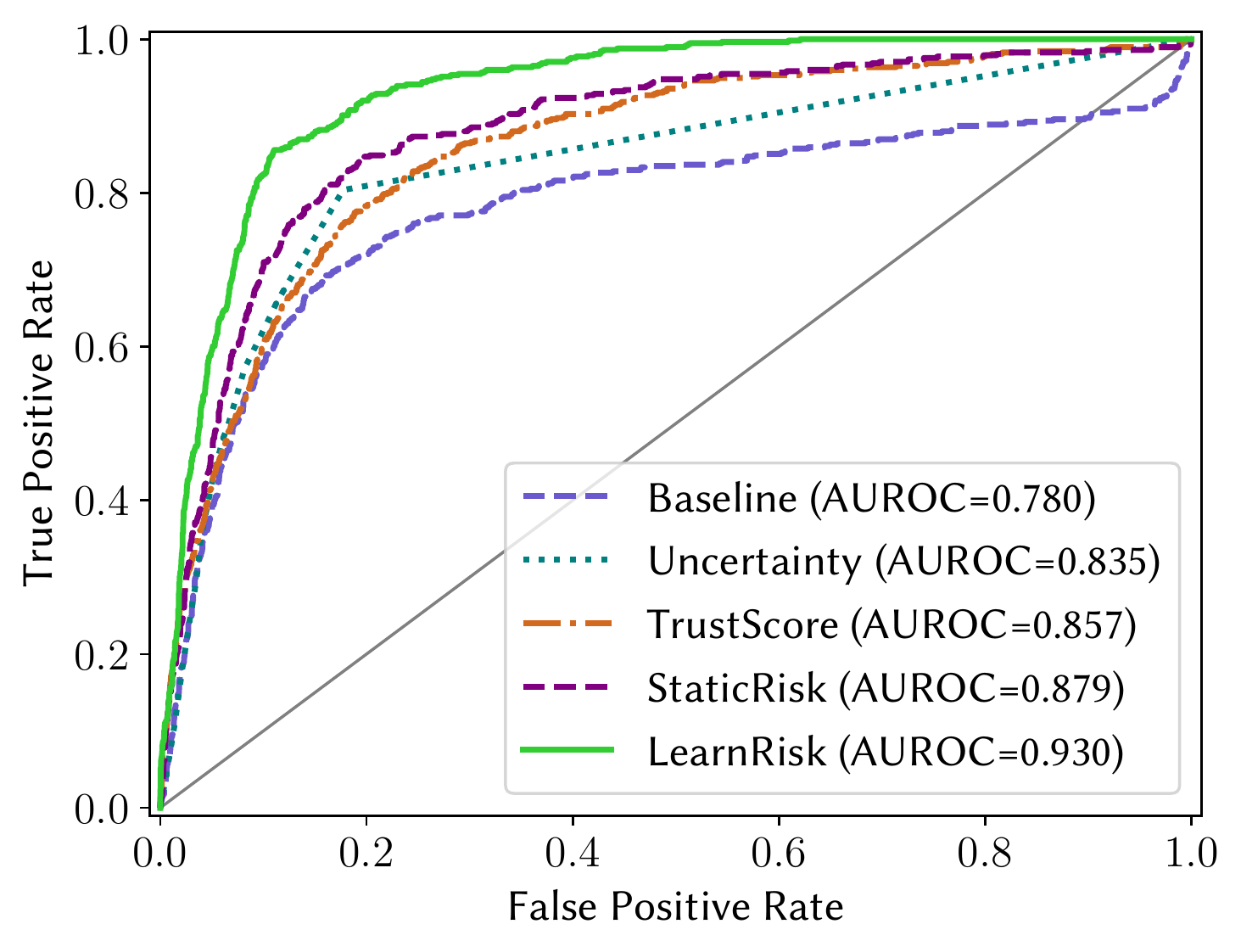}}
	\subfigure[SG(1:2:7)]
	{\includegraphics[width=\figw\linewidth]{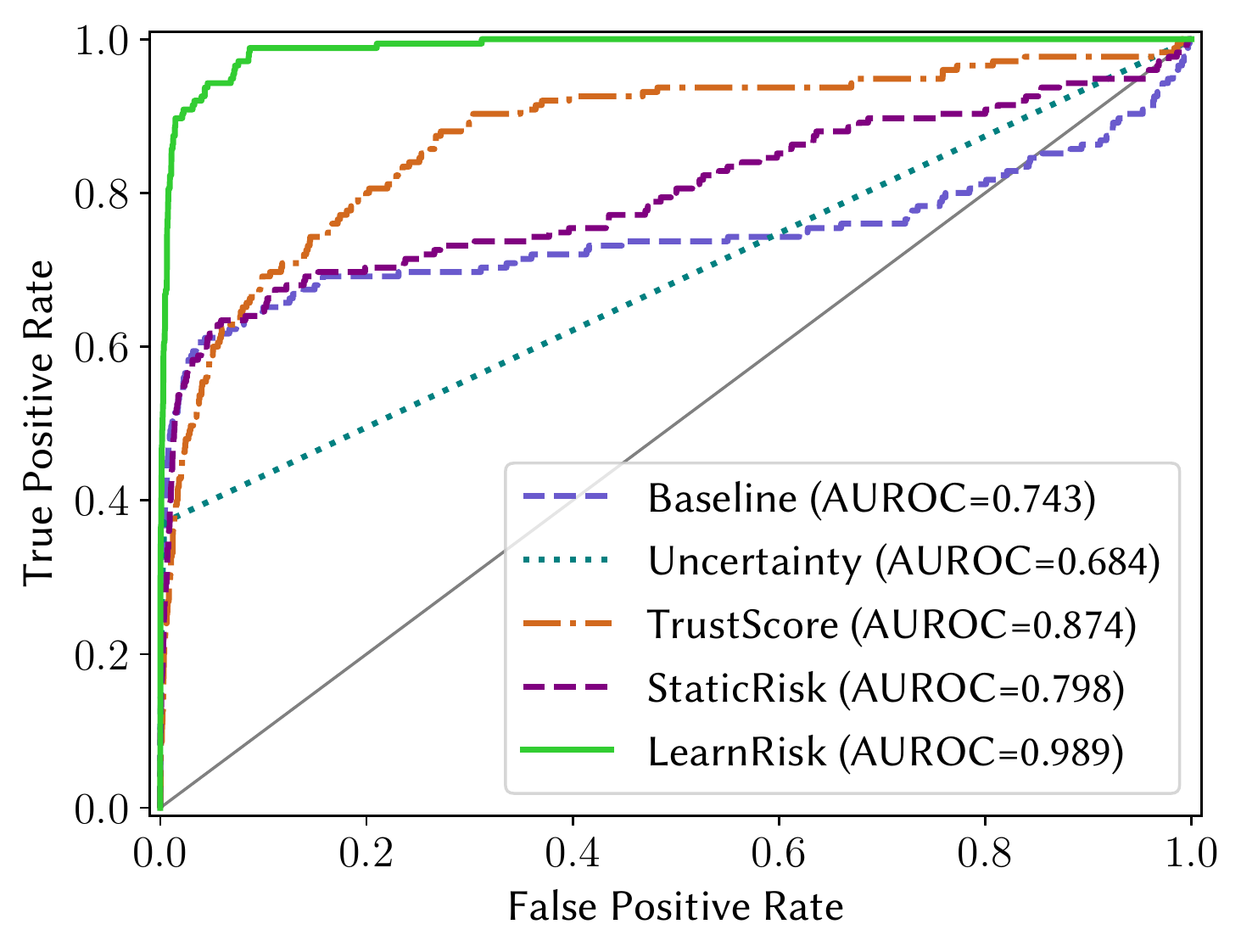}}
	\subfigure[SG(2:2:6)]
	{\includegraphics[width=\figw\linewidth]{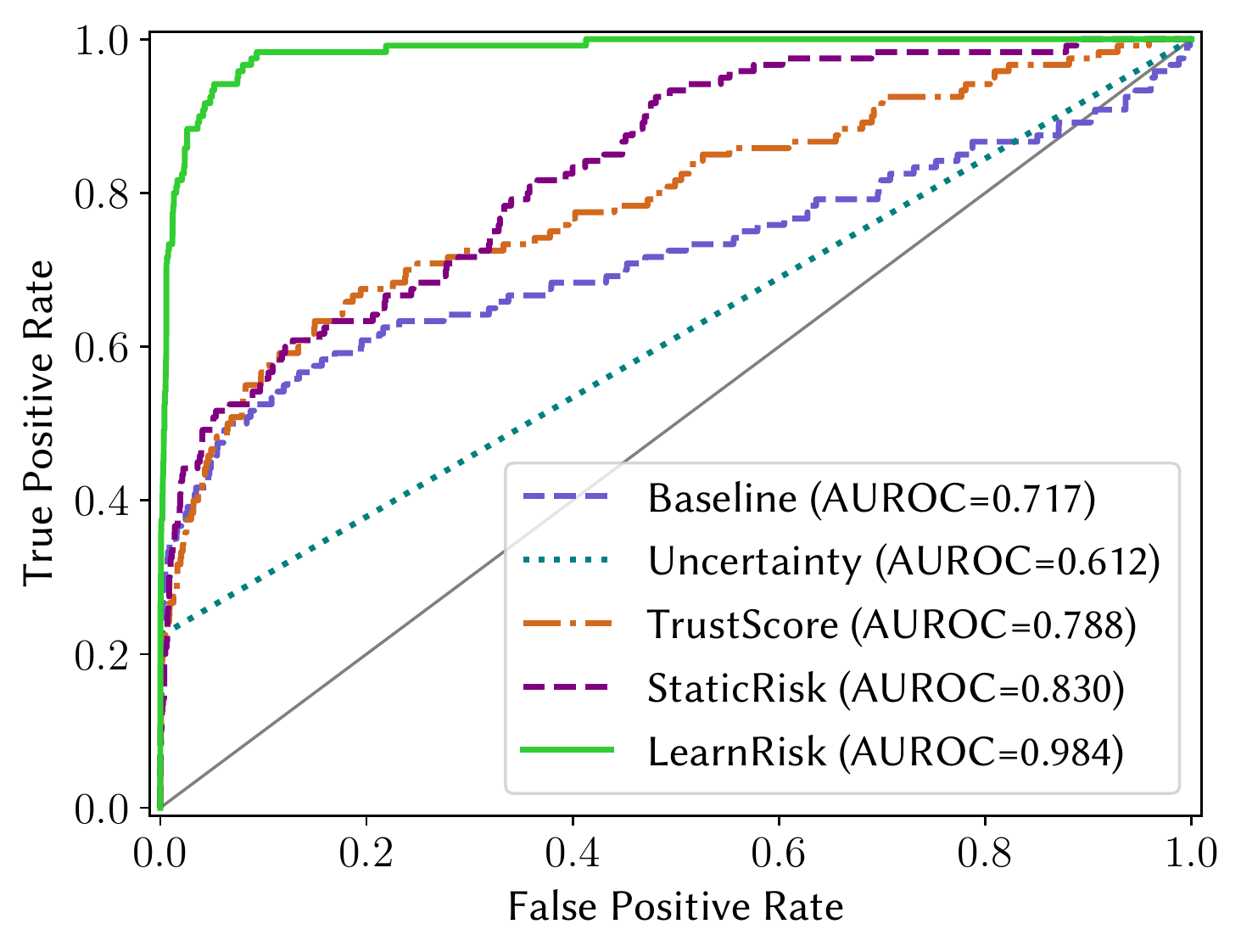}}
	\subfigure[SG(3:2:5)]
	{\includegraphics[width=\figw\linewidth]{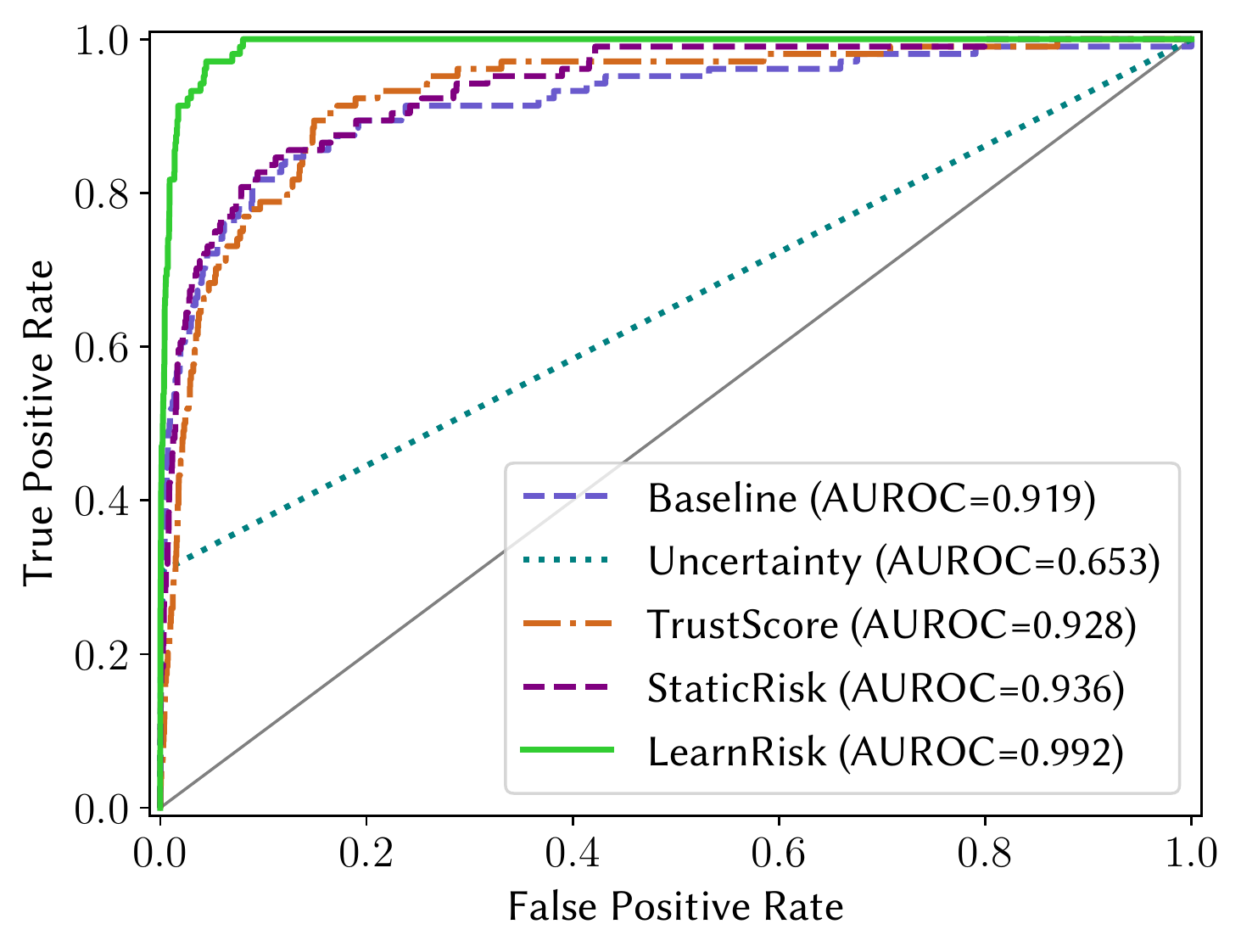}}
    \vspace{-0.15in}
	\caption{Comparative evaluation on four real datasets with varying ratios of (training:validation:test).}
	\vspace{-0.08in}
	\label{fig:experiments}
\end{figure*}

We have compared \emph{LearnRisk} to its alternatives in the experimental settings with various ratios of train, validation and test. The purpose is to track the accuracy of risk analysis under different circumstances with various machine classifier performance. 
The ratio has been set at 1:2:7, 2:2:6 and 3:2:5 respectively. 
The detailed evaluation results are presented in Figure~\ref{fig:experiments}. Note that \emph{Uncertainty} estimates the risk scores based on 20 trained models, which can only generate 21 different scores. Since many test data may share the same risk score, we have observed that \emph{Uncertainty} achieves highly regular ROC curves.

It can be observed that \emph{LearnRisk} consistently performs better than its alternatives under all the test circumstances. Specifically, \emph{LearnRisk} outperforms \emph{Baseline} and \emph{Uncertainty} by very considerable margins (more than 10\% measured by AUROC) in most cases. 
\emph{TrustScore} and \emph{StaticRisk} performs better than \emph{Baseline} and \emph{Uncertainty}. However, \emph{LearnRisk} still manages to outperform them by more than 5\% (measured by AUROC) in most cases. 
It is also worthy to point out that the performance of \emph{LearnRisk} is much more stable compared to its alternatives. For instance, on the SG workloads with different ratios, the achieved AUROCs of \emph{StaticRisk} are 0.798, 0.830 and 0.936 respectively, while the achieved AUROCs of \emph{LearnRisk} fluctuate only slightly (0.989, 0.984 and 0.992 respectively). Our experimental results have evidently shown that \emph{LearnRisk} performs better its alternatives in both accuracy and stability. They bode well for its efficacy in real scenarios. 

\vspace{0.1in}
\hspace{-0.15in}{\bf Out-of-distribution Evaluation.}  To further demonstrate the advantage of \emph{LearnRisk}, we have also evaluated the performance of various approaches in the out-of-distribution (OOD) circumstance, where the distribution of classifier training data is different from the validation and test data. This setting simulates the scenario where a pre-trained model is applied in a new environment. We have generated two OOD test workloads. In the first workload, the training data come from the DBLP-ACM dataset, while the validation and test data come from the DBLP-Scholar dataset. We denote this workload by DA2DS. In the second workload, the training data come from Abt-Buy, while the validation and test data come from Amazon-Google. As expected, our experiments showed that the performance of DeepMatcher on both OOD workloads deteriorates considerably compared to its performance on the DS and AB workloads.   

\begin{figure}
	\centering
	\subfigure[DA2DS dataset.]
	{\includegraphics[width=0.493\linewidth]{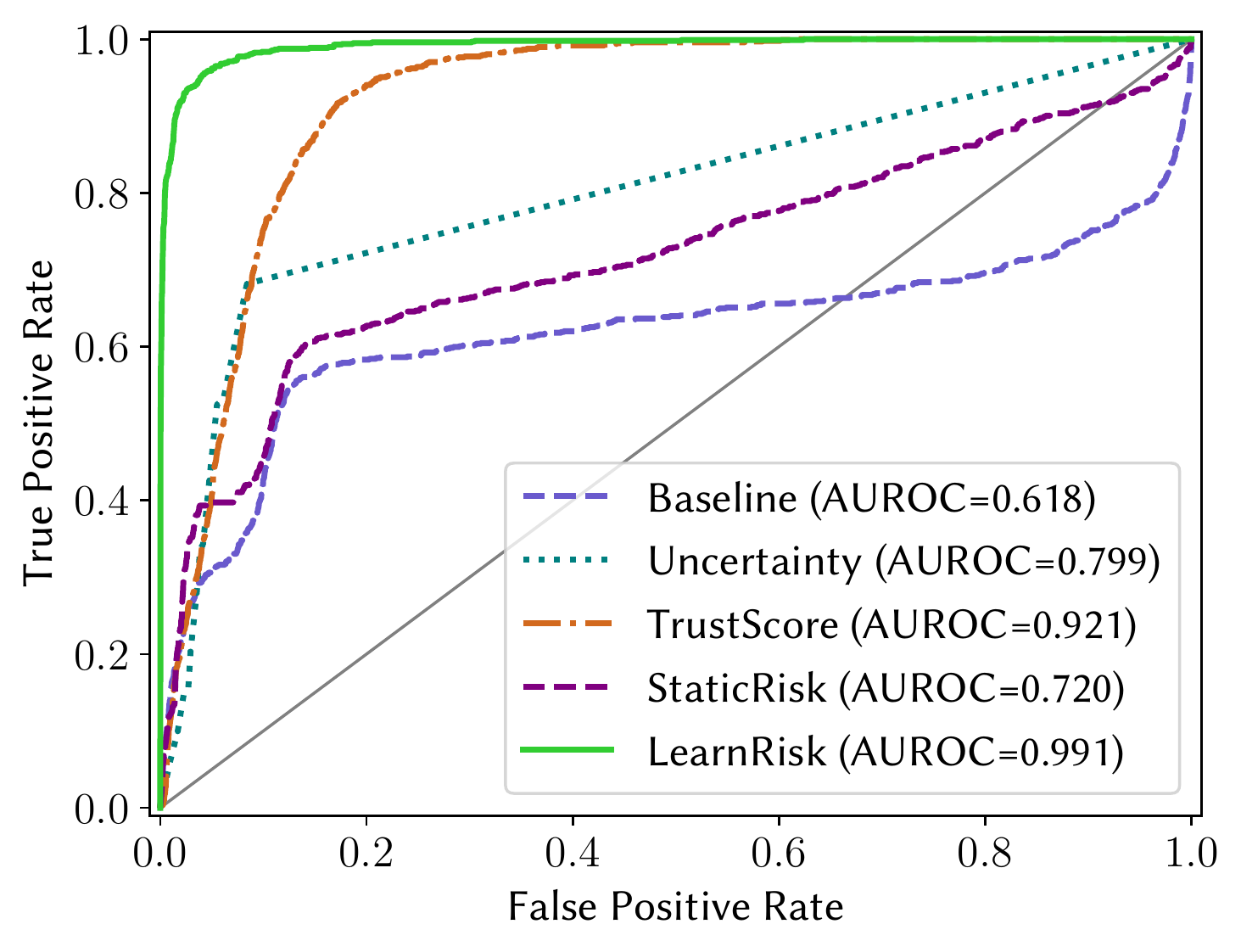}}
	\subfigure[AB2AG dataset.]
	{\includegraphics[width=0.493\linewidth]{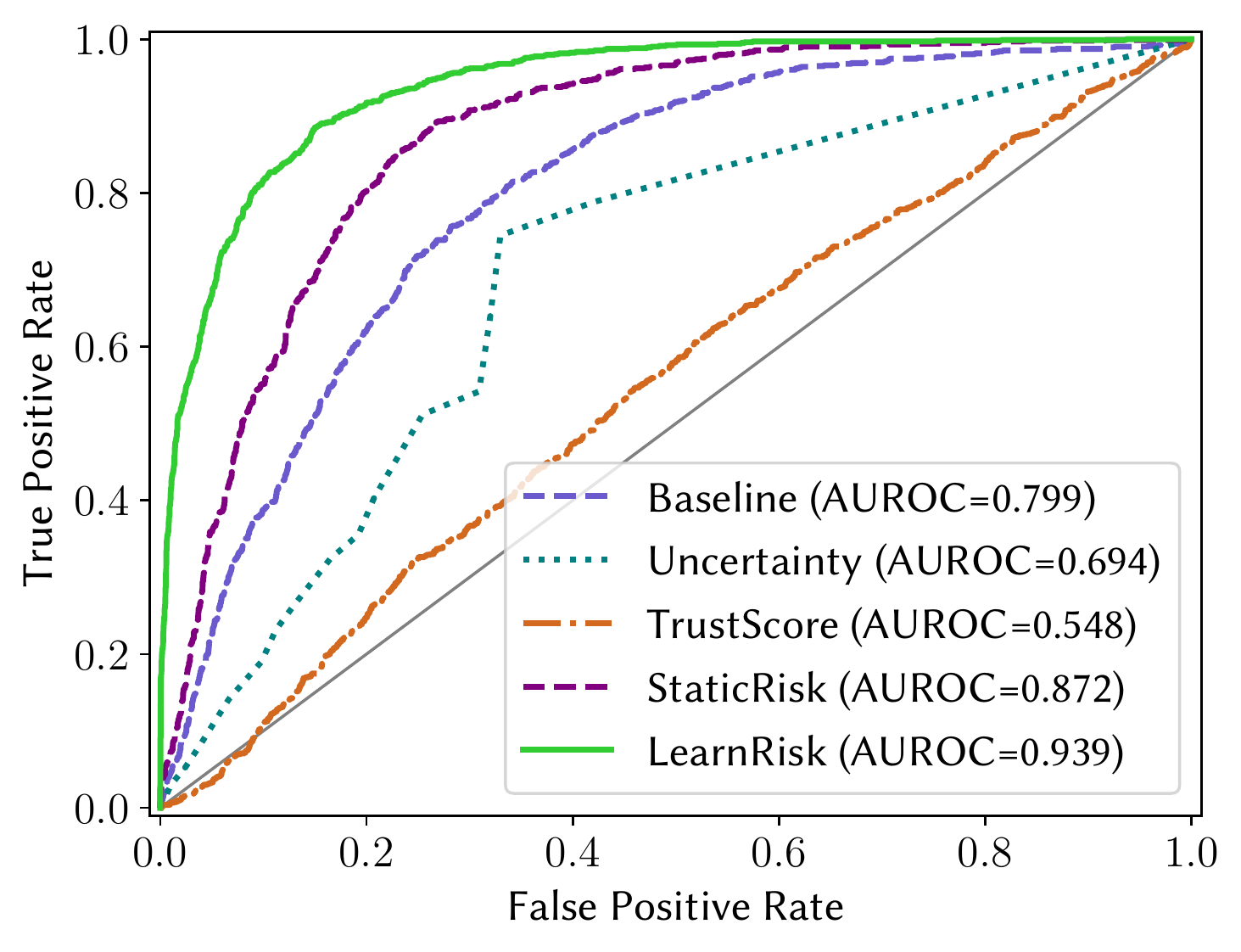}}
	\vspace{-0.15in}
	\caption{Comparative evaluation on the out-of-distribution datasets.}
	\label{fig:experiments-3}
\end{figure}

The detailed comparative evaluation results are presented in Figure \ref{fig:experiments-3}. It can be observed that similar to the same-distribution case, \emph{LearnRisk} consistently performs better than its alternatives on all the test cases. As expected, the performance margins between \emph{LearnRisk} and its alternatives become more considerable. It is also interesting to point out that the performance of the non-learnable alternative risk models fluctuates wildly between the two OOD workloads. For instance, \emph{TrustScore} performs well on DA2DS with the AUROC score of 0.921, while its performance on AB2AG is much worse with the AUROC score of 0.548. It can also be observed that \emph{StaticRisk} performs much better on AB2AG than on DA2DS (0.872 vs 0.720). In comparison, the performance of \emph{LearnRisk} is much more stable with the scores of 0.991 and 0.939 on DA2DS and AB2AG respectively. These experimental results demonstrate that the proposed risk model can effectively learn the characteristics of a target workload for improved risk measurement.

\begin{figure*}
	\centering
	\subfigure[DS.]
	{\includegraphics[width=0.24\linewidth]{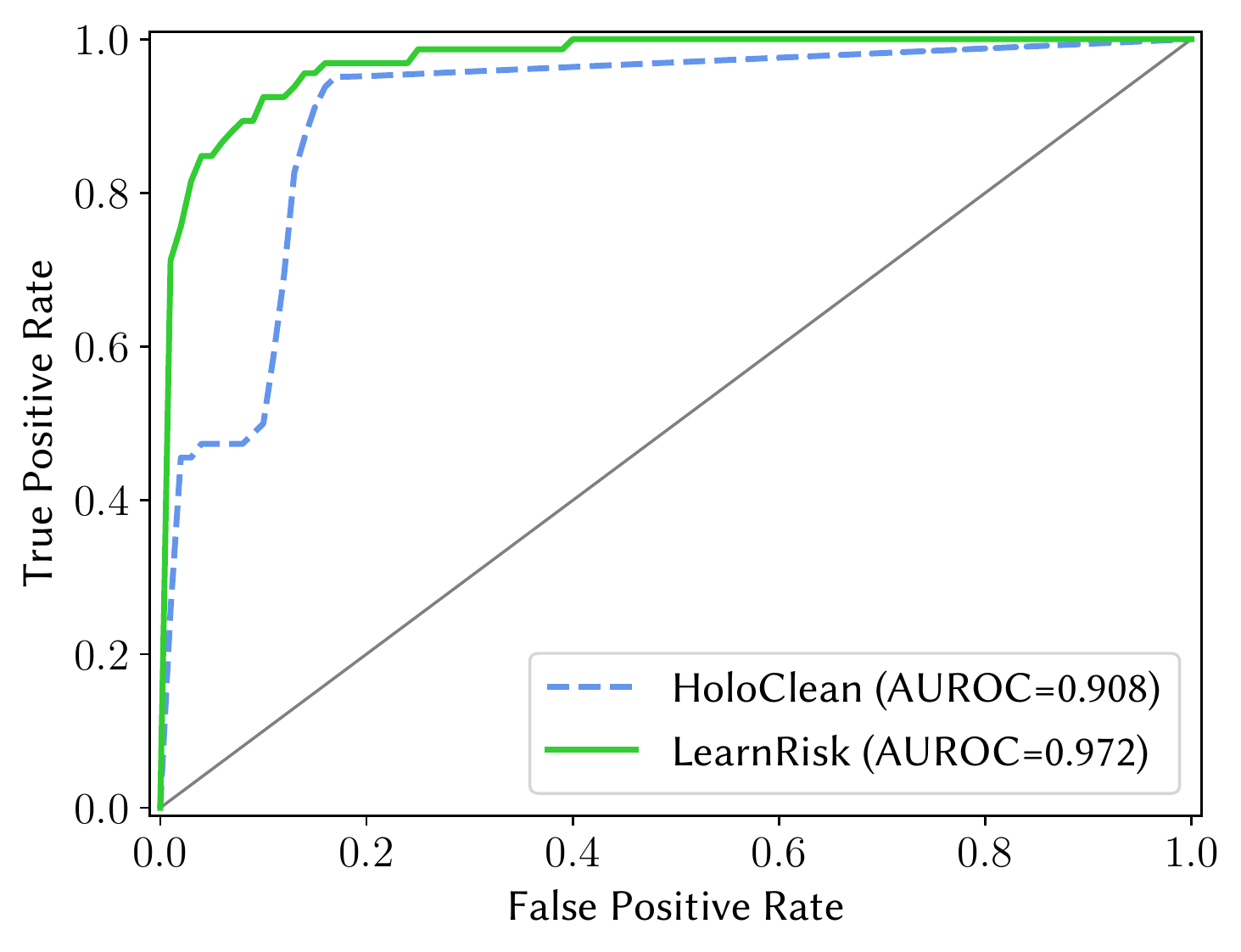}}
	\subfigure[AB.]
	{\includegraphics[width=0.24\linewidth]{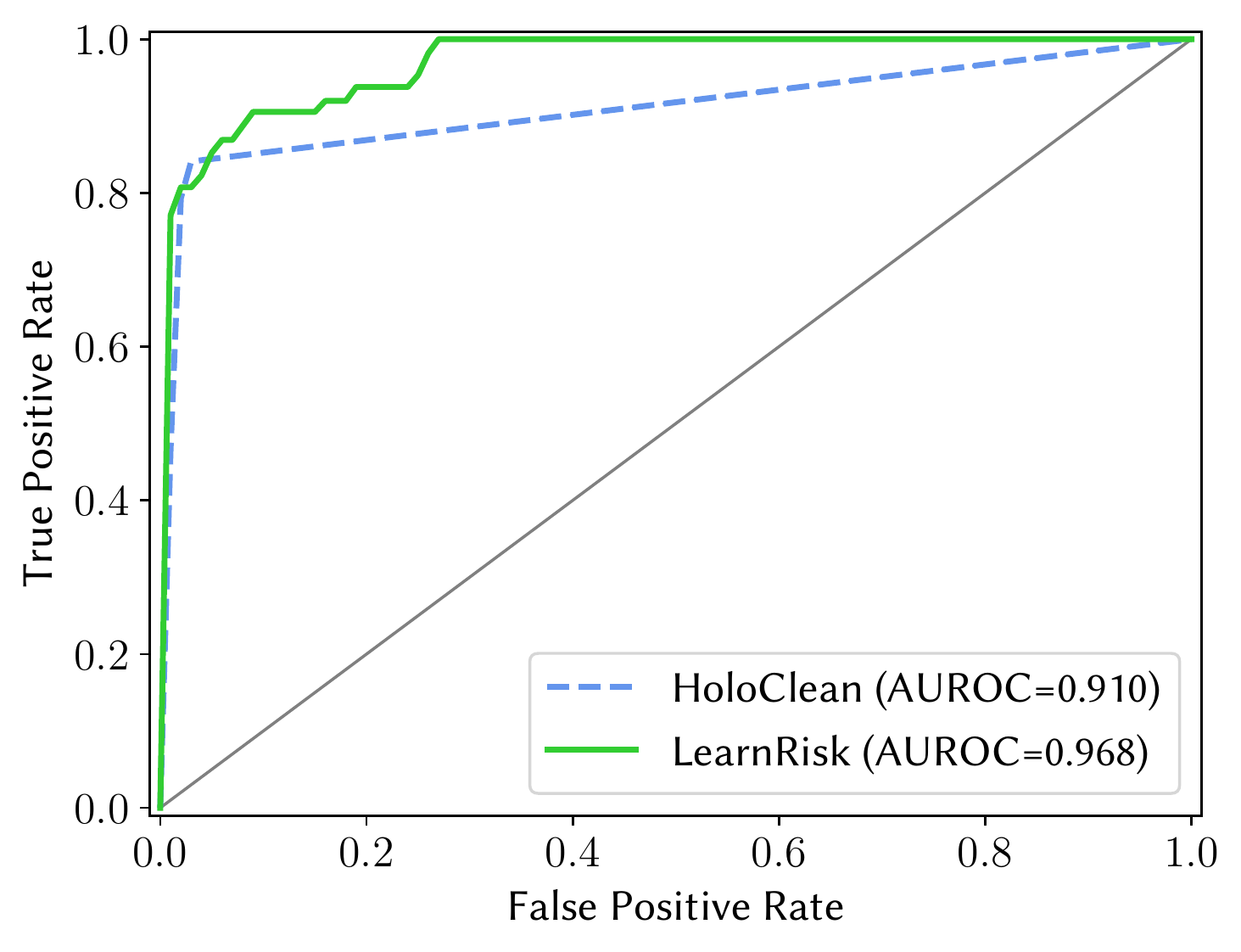}}
		\subfigure[AG.]
	{\includegraphics[width=0.24\linewidth]{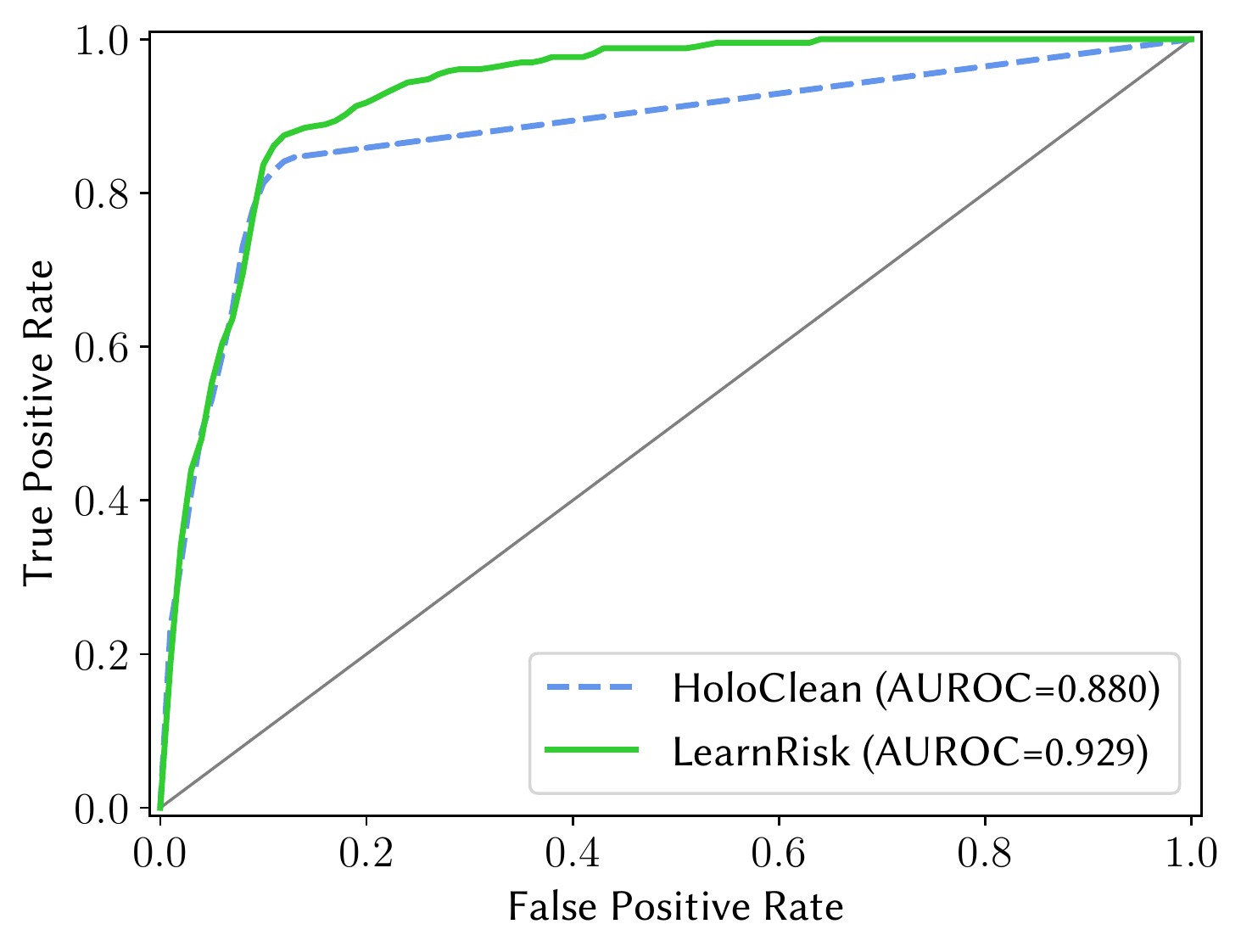}}
		\subfigure[SG.]
	{\includegraphics[width=0.24\linewidth]{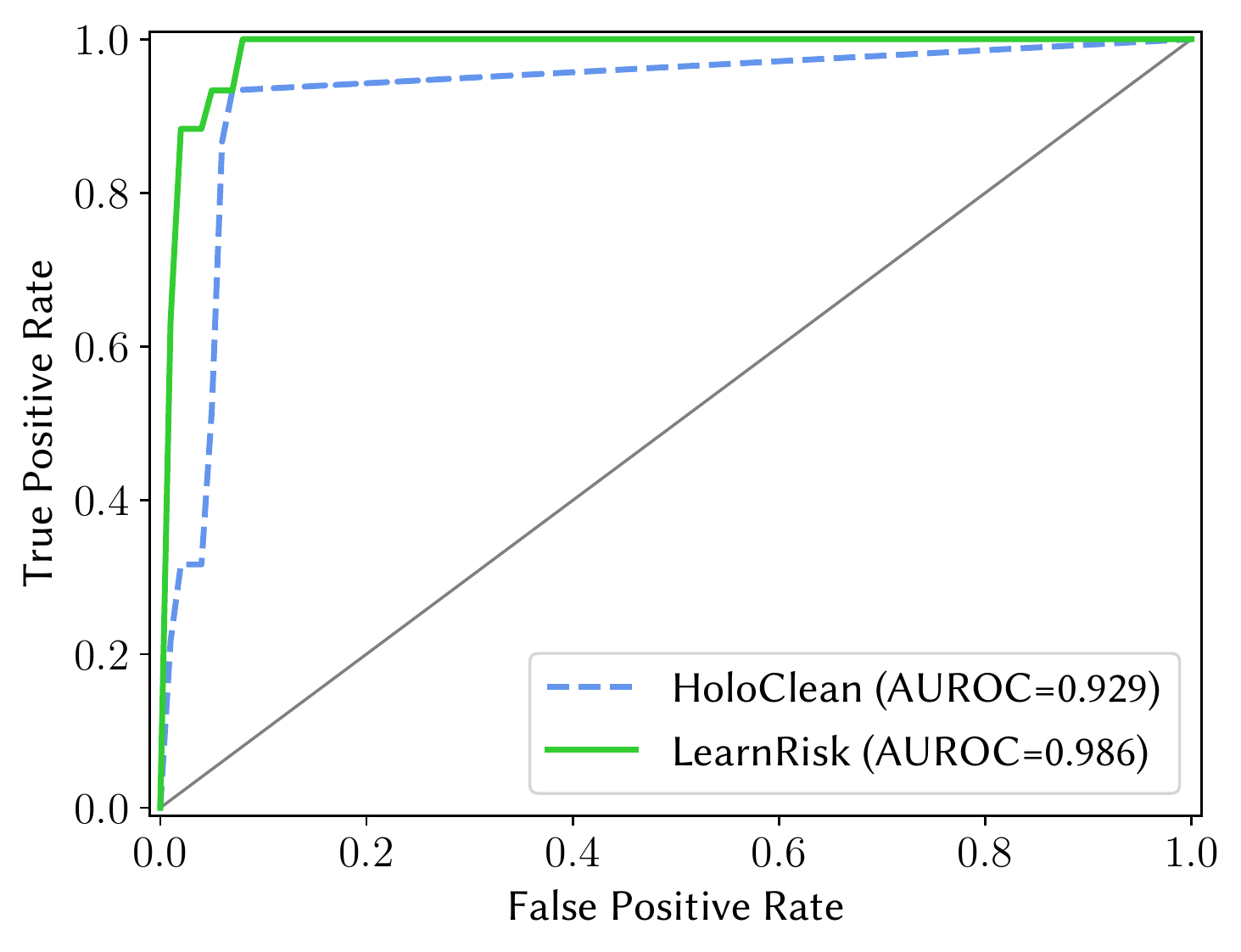}}
	\vspace{-0.15in}
	\caption{Comparison with HoloClean.}
	\vspace{-0.1in}
	\label{fig:holo-comparison}
\end{figure*}

\begin{figure*}
	\centering
	\subfigure[DS (Random).]
	{\includegraphics[width=\figw\linewidth]{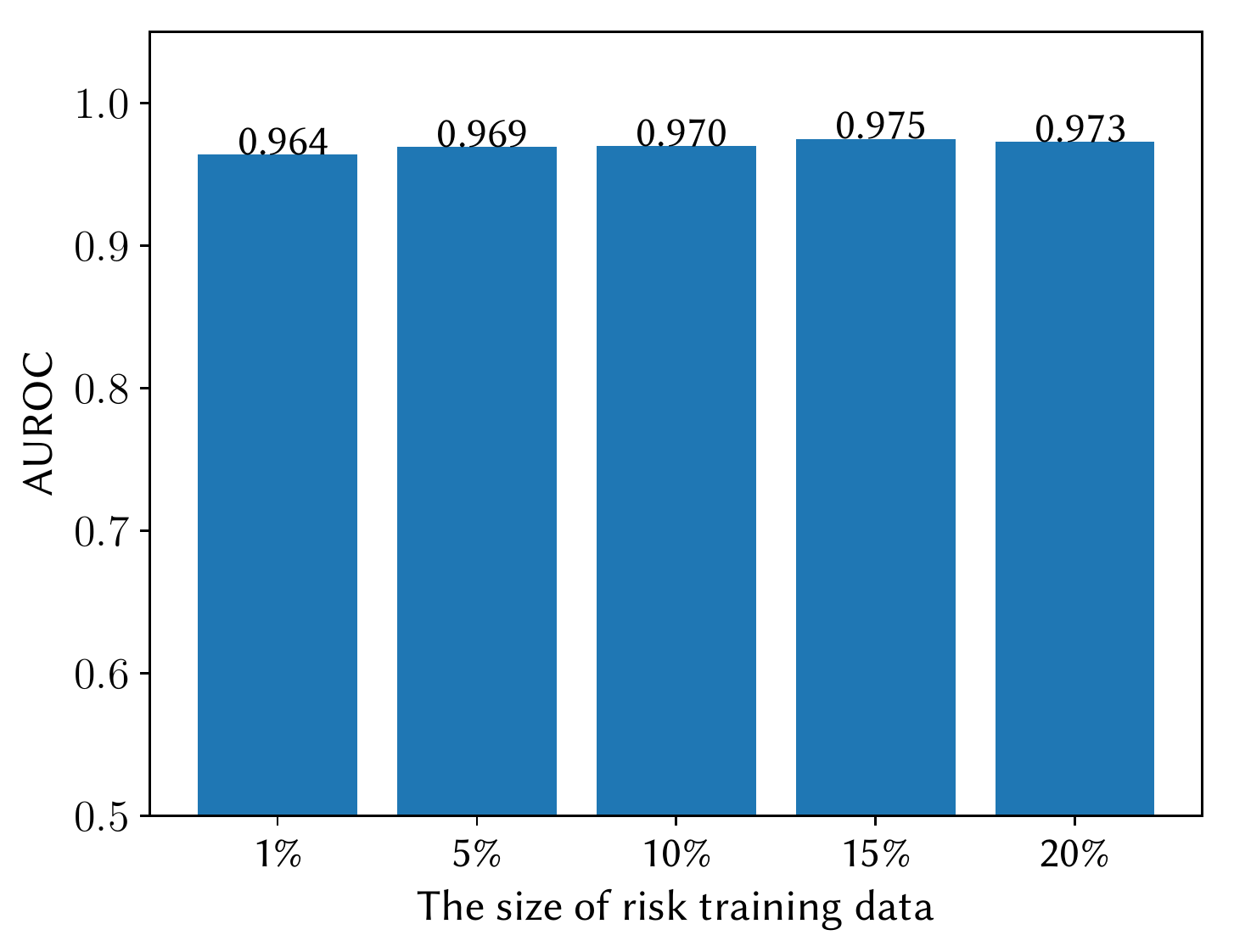}}
	\subfigure[AB (Random).]
	{\includegraphics[width=\figw\linewidth]{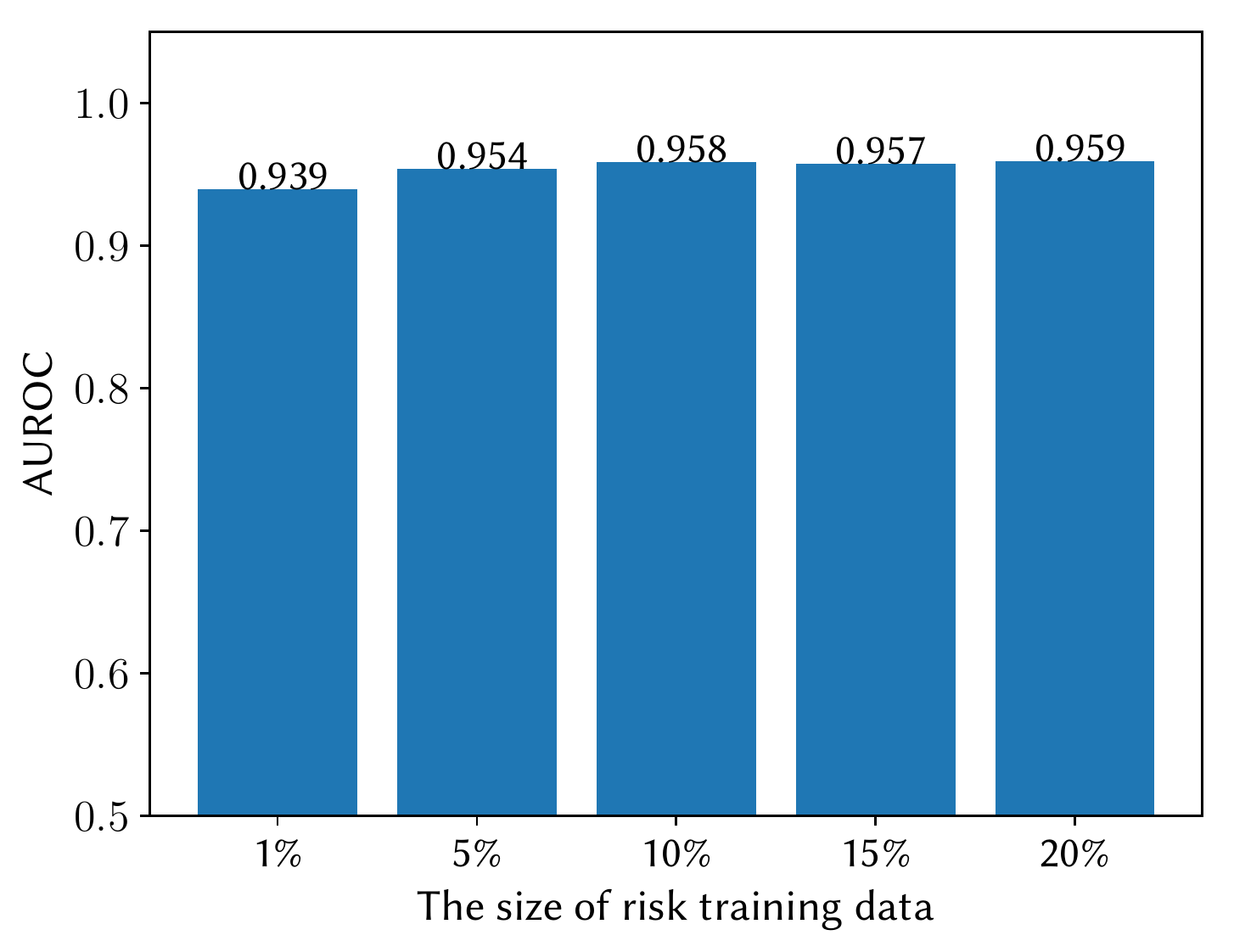}}
	\subfigure[DS (Active).]
	{\includegraphics[width=\figw\linewidth]{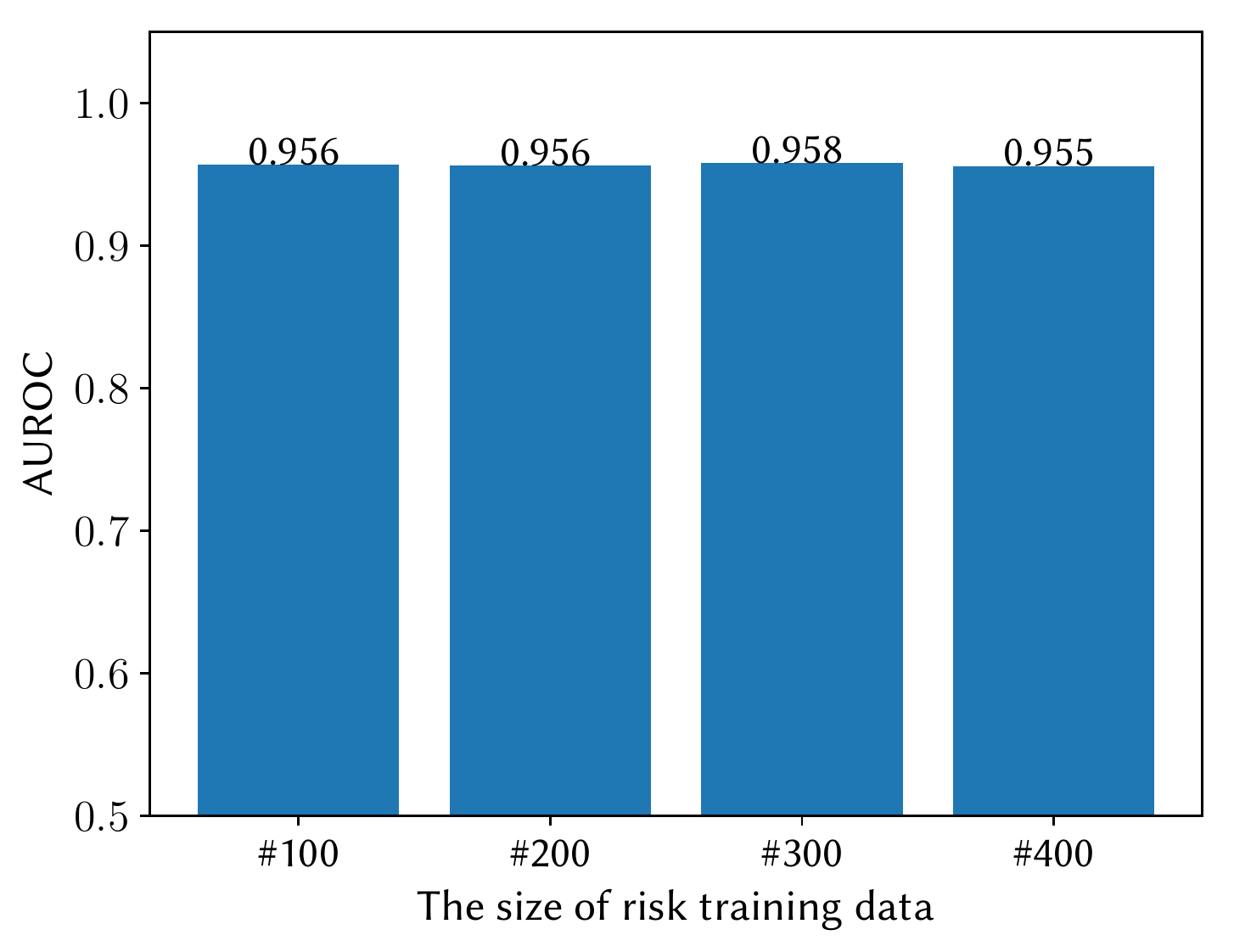}}
	\subfigure[AB (Active).]
	{\includegraphics[width=\figw\linewidth]{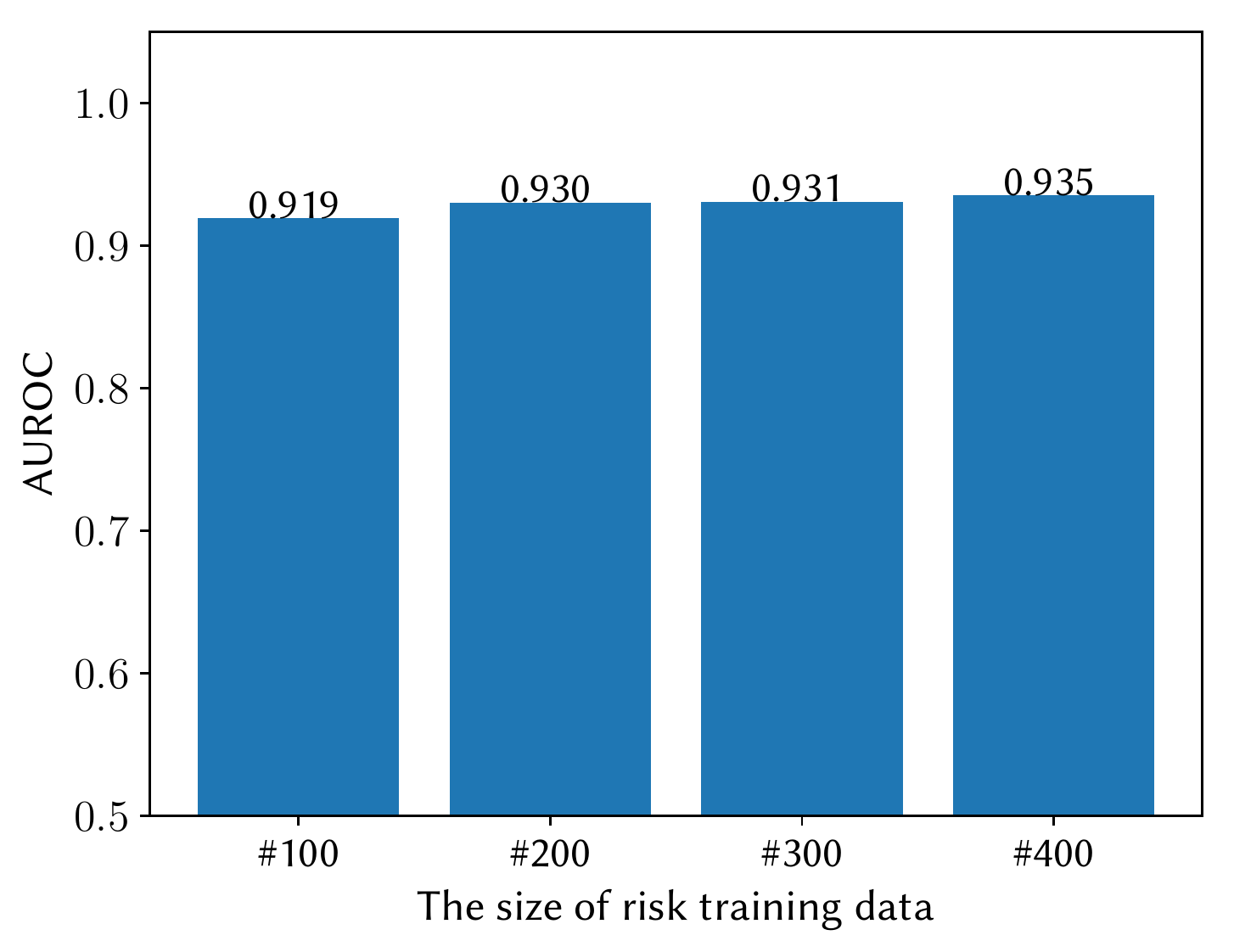}}
	\vspace{-0.15in}
	\caption{Performance Sensitivity Evaluation of LearnRisk w.r.t size of risk training data: in (a) and (b), the training data are selected by random sampling; in (c) and (d), they are selected by active learning.}
	\label{fig:experiments-sensitivity}
\end{figure*}

\subsection{Comparison with HoloClean}\label{sec:holoclean-comparison}

HoloClean~\cite{rekatsinas2017holoclean} is a data cleaning solution based on probabilistic inference, but can be adapted for ER. For risk analysis, we treat a candidate pair as a tuple in a relational table, with the labeling rules as its attributes. Considering the rules as a set of integrity constraints over the data, \emph{HoloClean} infers the probabilities of suggested status for noisy machine labels, which can be used to indicate risk of being mislabeled. As in~\cite{gokhale2014corleone}, we apply random forest~\cite{breiman2001random} to generate the labeling rules. The input features for the random forest are the same basic metrics used by \emph{LearnRisk}, except that it additionally uses DNN output as one of the metrics. As in LearnRisk, we set the maximum tree depth to 4 and the minimum number of samples to 5. For each test case, the number of generated labeling rules is also set to be very close to the number of one-sided rules generated by LearnRisk. We used the open-sourced implementation of \emph{HoloClean}~\footnote{https://github.com/HoloClean/holoclean} to support risk analysis.

Note that effective ER risk analysis requires a large number of rules (e.g. hundreds in our experiments) to ensure high pair coverage. Unfortunately, with hundreds of rules, HoloClean does not scale well with test data size: for instance, with only 295 rules and 2000 pairs from the DS dataset, Holoclean runs more than 24 hours on our machine with four Xeon E7-4820 CPUs, 630GB RAM, running Centos 6.1. Therefore, for each dataset, we generate the test workloads consisting of 1000 pairs, which are randomly sampled from the original dataset; the only exception is SG, in which the test workloads consist of 2000 pairs, because 1000-pair workloads contain few mislabeled instances. For each dataset, we generate 5 subsets and report the average performance over them. Figure~\ref{fig:holo-comparison} shows the comparative evaluation results. We can see that \emph{LearnRisk} consistently performs better than \emph{HoloClean} on all test cases. Our close scrutiny reveals that aiming to correctly label pairs in both ways, the two-sided labeling rules usually have limited efficacy in identifying risky pairs.	

\subsection{Performance Sensitivity}\label{sec:sensitivity}

   In this subsection, we evaluate the performance sensitivity of the proposed risk model with regard to the size of risk training data. 
For sensitivity evaluation, we fix the classifier training data and the test data, but vary the size of risk training data. In our experiments, the percentages of classifier training and test data are set to be 30\% and 50\% of the original dataset respectively, but the size of risk training data is varied from 100 upward.

We evaluate performance sensitivity by two experiments. In the first experiment, we select the data points for risk training by random sampling. In the second experiment, we select the data points for risk training in an active learning manner. Specifically, we iteratively select the data points with the highest ambiguity score. The detailed evaluation results on DS and AB are presented in Figure~\ref{fig:experiments-sensitivity}. The experimental results on AG and SG are similar, thus omitted here due to space limit. It can be observed that with random sampling, the performance of \emph{LearnRisk} is very robust w.r.t the size of risk training data: in the wide arrange between 1\% and 20\%, the performance of \emph{LearnRisk} only fluctuates very slightly. Even with the percentage of risk training data at 1\%, \emph{LearnRisk} outperforms all its alternatives in accuracy. With active selection, we can observe that even with only 100 pairs as risk training data, the performance of \emph{LearnRisk} is better than all its alternatives on DS, and with only 200 training data, the performance of \emph{LearnRisk} is highly competitive with the best of its alternatives on AB. Our experimental results have evidently demonstrated that the risk model of \emph{LearnRisk} can be effectively learned based on only a few carefully selected labeled instances. They further bode well for its efficacy in real scenarios.

\begin{figure}
	\centering
	\subfigure[Runtime of rule generation.]
	{\includegraphics[width=0.48\linewidth]{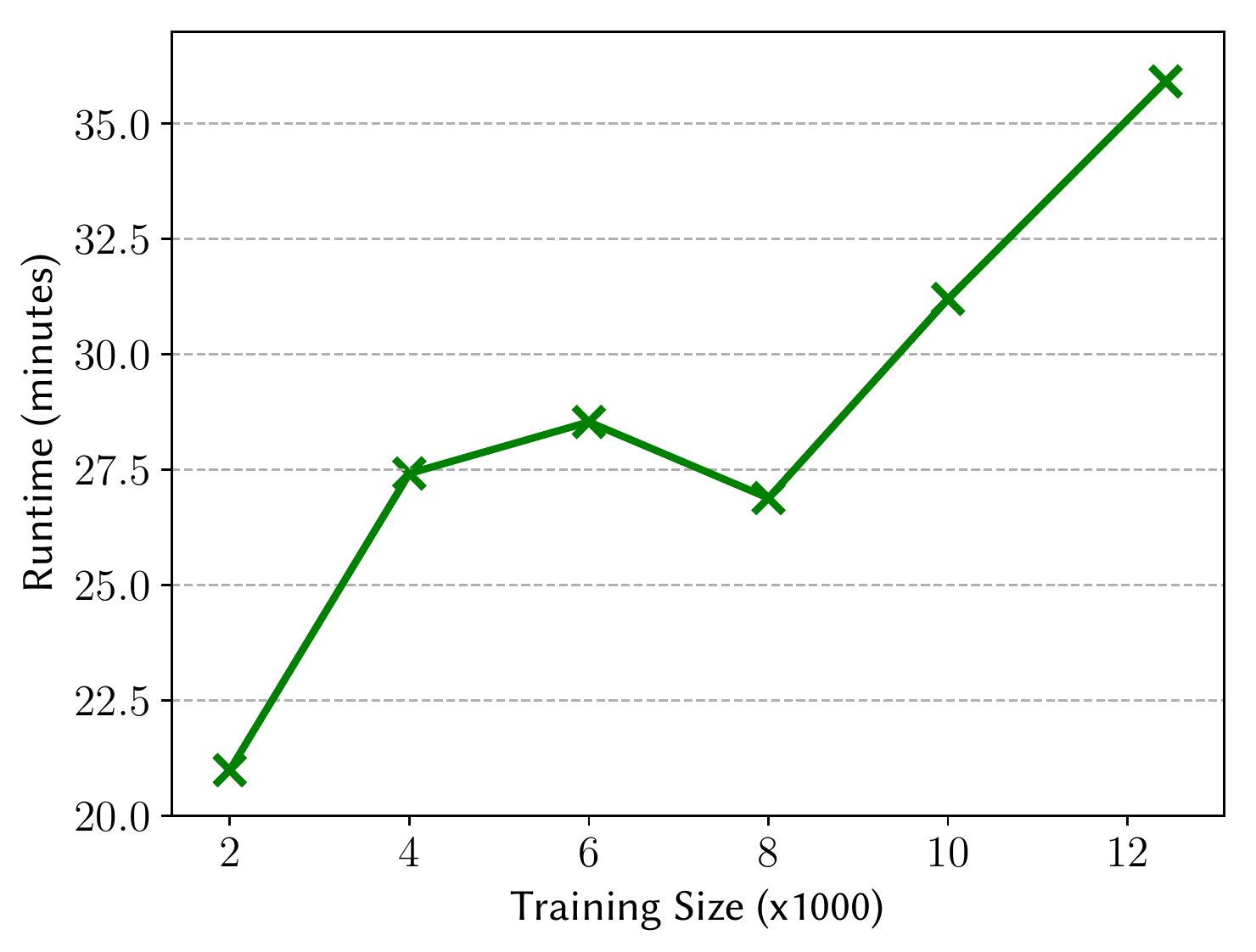}
	 \label{fig:DS-rule-scalability}}
	\subfigure[Runtime of \emph{LearnRisk}.]
	{\includegraphics[width=0.48\linewidth]{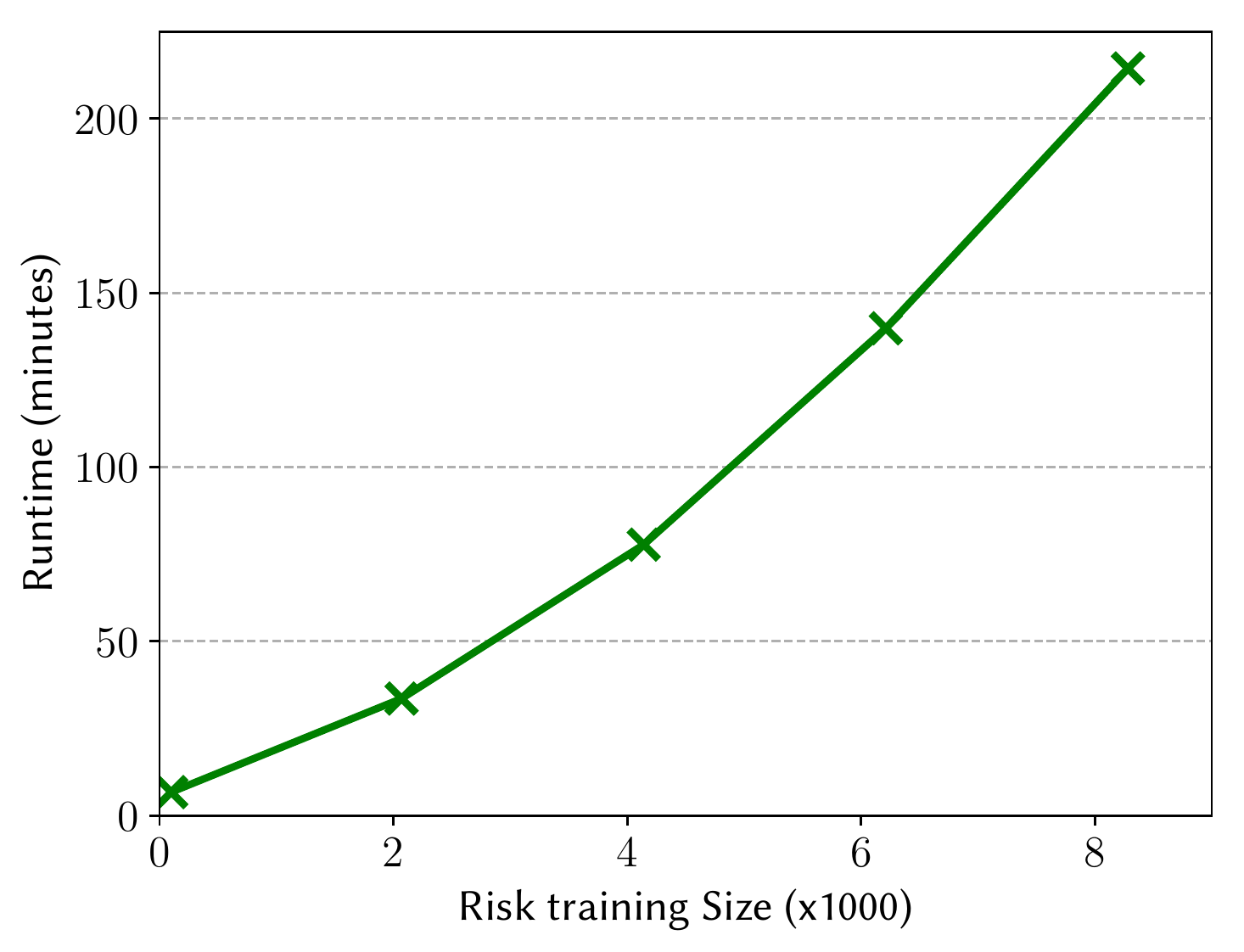}
	 \label{fig:DS-scalability}}
	 \vspace{-0.12in}
	\caption{Scalability evaluation on DS.}
	\vspace{-0.07in}
	\label{fig:experiments-scalability}
\end{figure}

\subsection{Scalability Evaluation}\label{sec:scalability} 

We first evaluate the scalability of risk feature generation. We have evaluated its scalability on various DS workloads. The detailed results are presented in Figure~\ref{fig:DS-rule-scalability}. The results on other datasets are similar, thus omitted here. It can be observed that the runtime consumed by rule generation generally increases linearly with the size of training data. Recall that in rule generation, the impurity of training data plays a key role in determining the number of rules generated. More training data do not necessarily result in more rules. Hence, there exist some fluctuations on runtime between different data sizes. Secondly, we evaluate the scalability of risk model training. Similarly, we have evaluated its scalability on the DS workloads. The detailed results are presented in Figure~\ref{fig:DS-scalability}. It can be observed that similarly, the runtime consumed by model training increases approximately linearly with the size of risk training data. Our experimental results have evidently shown that \emph{LearnRisk} scales well with workload.

%% file: 7-conclusion.tex
\section{Discussion on Potential Applications of Risk Analysis} \label{sec:potential-application}

\begin{figure}
	\centering
	\includegraphics[width=0.65\linewidth]{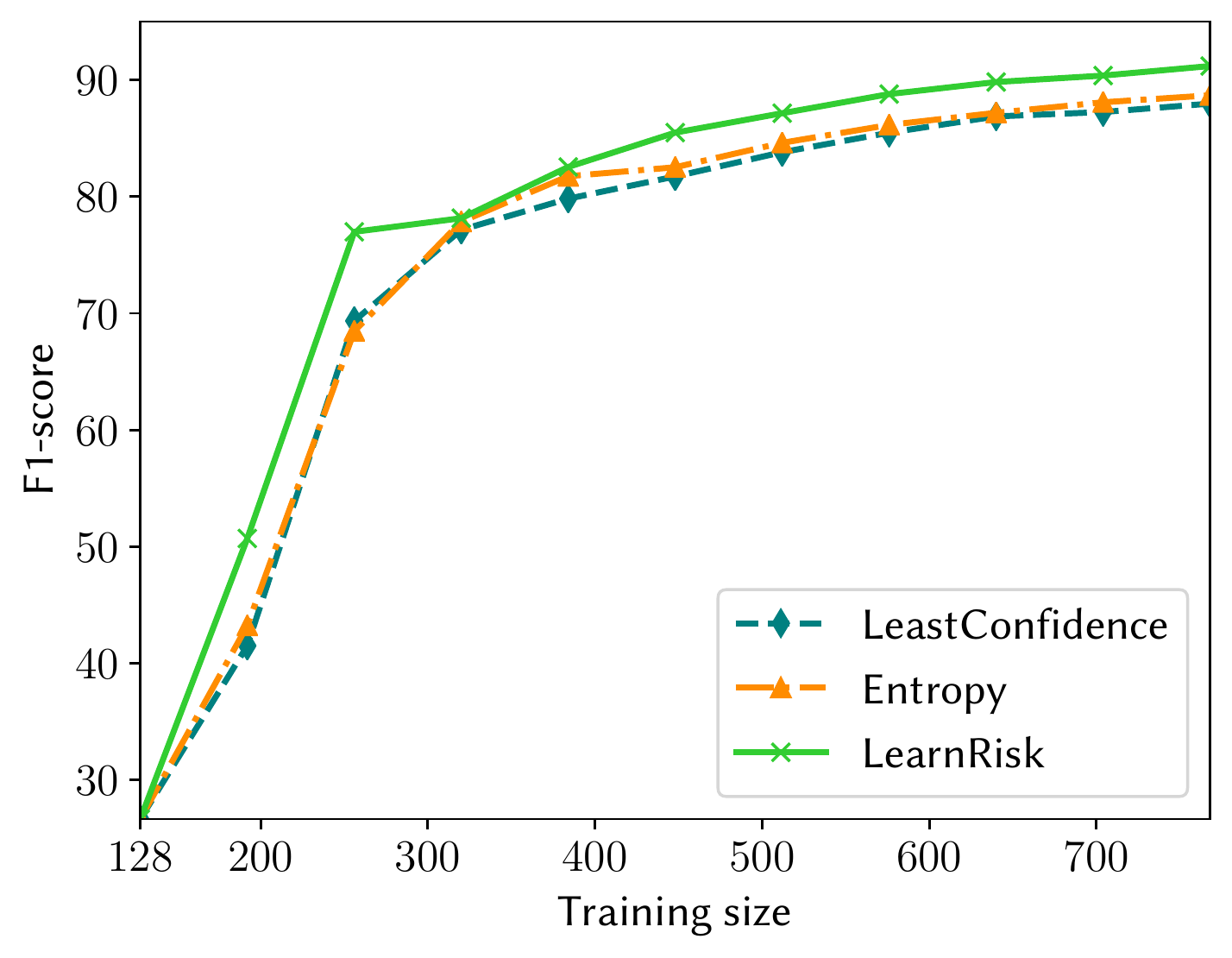}
	\vspace{-0.1in}
	\caption{ER Active Learning with \emph{LearnRisk}.}
	\vspace{-0.1in}
	\label{fig:DS-AL}
\end{figure}

It is worthy to point out that even though this paper considers risk analysis as a separate process independent of ER classifier training, risk analysis can be potentially leveraged to improve classifier training. We discuss two potential applications, active selection of training instances and model training:

\vspace{0.1in}
\hspace{-0.15in}{\bf Active Learning.} The existing techniques for active learning~\cite{huang2010active, wang2015querying} are mainly based on the metric of uncertainty, or the hybrid metrics combining uncertainty and other dimensions (e.g. representativeness). It has been empirically shown~\cite{Daniel2019discriminative} that if the batch size of each iteration is set to be large (e.g. 1000), the uncertainty-based metric can usually achieve the overall best performance. Since our proposed approach for risk analysis can provide more accurate uncertainty measurement compared with the existing ones, it can be naturally leveraged for active selection of training instances. At each iteration, the algorithm can select the most risky instances for labeling.

We have conducted an experiment on the DS dataset to compare the proposed risk approach with the uncertainty-based approaches, \emph{LeastConfidence} and \emph{Entropy}. A DeepMatcher model is initially trained on $L$, then it is iteratively retrained as  more data is acquired in batches. With $|L|$=128 and the batch size being equal to 64, the evaluation results on the DS dataset are presented in Figure~\ref{fig:DS-AL}.
It can be observed that \emph{LearnRisk} performs better than both \emph{LeastConfidence} and \emph{Entropy}. Due to space limit, further exploration and evaluation are beyond the scope of this paper. However, this preliminary experiment demonstrates that \emph{LearnRisk} is a promising approach for ER active learning.

\vspace{0.1in}
\hspace{-0.15in}{\bf Model Training.} The existing approaches for classifier training usually learn a model whose predictions on the training instances are most consistent with their ground-truth labels. However, the resulting classifier may not perform well on a target workload due to distribution misalignment. Since our proposed approach can provide reliable risk analysis on the classifier labels of target instances, the potential improved approach for classifier training is to train a model on both the labeled training instances and the unlabeled target instances. On the unlabeled instances, the objective is to minimize the prediction risk of classifier labels. Specifically, instead of only considering prediction accuracy on the labeled training instances, the revised objective function for model training consists of two parts, one to enforce the label consistency on the labeled training instances, the other to minimize the prediction risk on the unlabeled target instances.

\vspace{0.15in}
\section{Conclusion} \label{sec:conclusion}

In this paper, we have proposed an interpretable and learnable solution for ER risk analysis. Our extensive experiments on real data have validated its efficacy. Our solution focuses on ER, but the proposed framework can be potentially generalized to other classification tasks. Moreover, the risk approach can be potentially leveraged to improve ER classifier training, thus open an interesting research direction.